\documentclass[11pt]{article}

\usepackage{amsmath,amsfonts,amsbsy,amssymb,array,accents,dsfont}
\usepackage{enumerate,array,latexsym,graphicx,mathrsfs,verbatim,psfrag}
\usepackage{bm} 
\usepackage[normalem]{ulem}
\usepackage{booktabs} 
\usepackage[usenames]{color}
\usepackage[utf8]{inputenc}
\usepackage[english]{babel}
\usepackage{fancybox}

\usepackage{datetime}

\usepackage[nosort]{cite}
\usepackage{chngpage} 

\usepackage{enumitem}
\setlist{itemsep=0pt}

\usepackage[colorlinks=true,      linkcolor=darkblue,      urlcolor=darkblue,      
            filecolor=darkblue,      citecolor=darkblue,       pdfstartview=FitH,     
						pdfpagemode=UseNone,      bookmarksopen=true]{hyperref}  
\usepackage[all]{hypcap}     

\newcommand{\captionfonts}{\small}
\makeatletter  
\long\def\@makecaption#1#2{%
  \vskip\abovecaptionskip
  \sbox\@tempboxa{{\captionfonts #1: #2}}%
 \ifdim \wd\@tempboxa >\hsize
    {\captionfonts #1: #2\par}
  \else
    \hbox to\hsize{\hfil\box\@tempboxa\hfil}%
  \fi
  \vskip\belowcaptionskip}
\makeatother   

\DeclareMathSymbol{\medhatsym}{\mathord}{largesymbols}{"62} 

\DeclareMathSymbol{\medtildesym}{\mathord}{largesymbols}{"65}

\newcommand{\comm}[1]{} 

\def\IC{\mathbb{C}}

\def\IT{\mathbb{T}}

\def\({\left(}
\def\){\right)}
\def\[{\left[}
\def\]{\right]}

\def\sst{\scriptscriptstyle}

\def\coeff#1#2{{\textstyle \frac{#1}{#2}}}

\def\One{{\hbox{ 1\kern-.8mm l}}}

\def\barray{\begin{array}}
\def\earray{\end{array}}
\def\be{\begin{equation}}
\def\ee{\end{equation}}
\def\bea{\begin{eqnarray}}
\def\eea{\end{eqnarray}}
\def\bal{\begin{align}}
\def\eal{\end{align}}

\def\\aff{mu}

\numberwithin{equation}{section} 

\makeatletter
\g@addto@macro\bfseries{\boldmath}
\makeatother

\definecolor{cardinal}{rgb}{0.6,0,0}
\definecolor{darkgreen}{rgb}{0,0.4,0}
\definecolor{purple}{rgb}{0.5, 0, 0.5}
\definecolor{golden}{rgb}{0.92, 0.7, 0}
\definecolor{midnight}{rgb}{0, 0, 0.5}
\definecolor{darkblue}{rgb}{0, 0, 0.8}

\def\IC{\mathbb{C}}

\def\coeff#1#2{\relax{\textstyle {#1 \over #2}}\displaystyle}

\def\ZZ{\mathds{Z}}

\def\cA{{\cal A}}

\def\cK{{\cal K}}

\def\cO{{\cal O}}

\def\aff{\mu}

\def\cO{{\cal O}}

\def\sst#1{\scriptscriptstyle{#1}}

\newcommand{\np}{\ensuremath{n_{\mathrm{\sst{P}}}}}

\begin{document}


\begin{flushright}
\end{flushright}

\vspace{14mm}

\begin{center}

{\huge \bf{Inscribing geodesic circles on the face of the superstratum}} \medskip \\


\vspace{13mm}

\centerline{{\bf  Bin Guo$^1$, Shaun D. Hampton$^2$  and Nicholas P. Warner$^{1,3,4}$}}
\bigskip
\bigskip
\vspace{1mm}

\centerline{$^1$ Institut de Physique Th\'eorique,}
\centerline{Universit\'e Paris Saclay, CEA, CNRS, }
\centerline{Orme des Merisiers,  F-91191 Gif sur Yvette, France}
\bigskip
\centerline{$^2$ School of Physics}
\centerline{Korea Institute for Advanced Study,}
\centerline{Seoul, 02455, Korea.}
\bigskip
\centerline{$^3$\,Department of Physics and Astronomy,}
\centerline{University of Southern California,} \centerline{Los
Angeles, CA 90089-0484, USA}
\bigskip  
\centerline{$^4$\,Department of Mathematics,}
\centerline{University of Southern California,} \centerline{Los
Angeles, CA 90089, USA}

\vspace{4mm}

%

\vspace{8mm}
 
\textsc{Abstract}

\begin{adjustwidth}{10mm}{10mm} 
 %
\vspace{3mm}
\noindent

We use families of circular null geodesics as probes of a family of microstate geometries, known as $(1,0,n)$ superstrata. These geometries carry a left-moving momentum wave and the behavior of some of the geodesic probes is very sensitive to this background wave.  The left-moving geodesics behave like  BPS particles and so can be placed in circular orbits anywhere in the geometry and actually ``float'' at fixed radius and angle in the three-dimensional ``capped BTZ'' geometry. The right-moving geodesics behave like non-BPS particles.  We show that they provide a simple geometric characterization of the black-hole bound: when the momentum charge of the geometry is below this bound, such geodesics can be placed anywhere, but  exceeding the bound, even by a small amount, means these geodesics are restricted to the deep interior of the  geometry.  We also  show that for  left-moving  string probes, the tidal forces remain comparable with those of global AdS$_3$.  Nevertheless, for some of these probes, the ``bumps'' in the geometry  induce an oscillatory mass term  and we discuss how this can lead to chaotic scrambling of the state of the string.

%
\end{adjustwidth}

\end{center}


\thispagestyle{empty}

\newpage


\baselineskip=14pt
\parskip=2pt

\tableofcontents


\baselineskip=15pt
\parskip=3pt

\section{Introduction}
\label{Sect:introduction}

Geodesics are  one of the simplest and most fundamental probes of geometries.  Moreover, because of the geometric optics approximation, they also provide information about possible solutions of the wave equation.  The equations of geodesic deviation then provide a  deeper insight into tidal stresses experienced by probes and the scattering of particles and waves moving through geometries.    In string theory, we can go one step further and examine the dynamics of strings as they move through diverse backgrounds, and one of the simplest first steps in this direction is to examine the Penrose   \cite{Penrose:1972ui,Penrose:1976xxxx}, or Penrose-G\"uven limit \cite{Gueven:2000ru} of such probes.

A string, or a particle, traveling at ultra-relativistic speeds only samples the immediate vicinity of the center of mass trajectory, and this trajectory is well approximated by a null geodesic. One can take a pencil of null geodesics around the original geodesic, and, at leading non-trivial order, the metric in this pencil becomes that of a plane wave. G\"uven generalized this to other families of background fields \cite{Gueven:2000ru}. A classical string can be solved and quantized in light-cone gauge in such a plane-wave background~\cite{Horowitz:1990sr}, and the  world-sheet dynamics reduces to that of a free field  in which the background fields create  a time-dependent mass matrix for the string excitations.   

Such stringy analyses of supergravity backgrounds became something of an industry 20 years ago, and, amongst other things, it led to an invaluable streamlining of the procedure of taking the Penrose limit and putting it in the Brinkmann form in which the string is readily solved.  A  review of the early technology can be found in \cite{Blau:2002dy, Blau:2002mw} and its streamlined version is discussed in \cite{Blau:2003dz,Blau:2004yi,Blau:2006ar}.  This simplified version relates the mass matrix felt by the string to an analogue of the geodesic deviation matrix for null geodesics.

In recent years there has been something of a resurgence in using geodesic deviation and stringy probes in the microstate geometry programme.  Microstate geometries closely approximate their black-hole counterparts until one is at the horizon scale: they are smooth and  horizon-free, and cap off smoothly  just above where the horizon would be in the black-hole solution.   The  throats of microstate geometries differ infinitesimally from those of black holes through multipole moments created by the cap geometry.      Despite these tiny differences,   it was shown in \cite{Tyukov:2017uig, Bena:2018mpb, Bena:2020iyw} that the ultra-relativistic speeds of infalling probes can magnify the multipole moments in such a manner that the tidal stresses reach the string scale before the probe even reaches the cap.   This led to the study of stringy probes \cite{Martinec:2020cml,Ceplak:2021kgl}, where it was shown that these tidal forces would excite strings significantly above their ground states, taking energy out of the  center of mass motion.  The result was ``tidal trapping:'' even massless infalling  string probes would be  excited and become trapped and scrambled into the microstate geometry.  Microstate geometries thus exhibited another of the defining features of a black hole. 

There are two motivations behind this paper:  to examine  classes of geodesics that may give rise to trapping and scrambling, and to look at the dynamics of strings that follow some of these geodesics.  Indeed, it was suggested several years ago that long-term trapping of time-like geodesics could lead to instabilities of microstate geometries to the formation of black holes or black rings \cite{Eperon:2016cdd,Chakrabarty:2019ujg}.     However, it has now been shown that the instability arising from long-term trapping  only occurs at sub-Planckian wavelengths \cite{Marolf:2016nwu, Bena:2020yii}, and is therefore an intrinsically stringy phenomenon.  Moreover, as suggested in  \cite{Bena:2018mpb}, it seems that a coherent expression of such an instability will simply cause a microstate geometry to evolve along its vast moduli space. 

The time-like geodesics that exhibit this extreme long-term trapping are those that limit to a particular closed null geodesic associated with an evanescent ergosphere. Indeed, this null geodesic lies at the heart of the microstate geometry. From a holographic perspective, this null geodesic is the original locus of the supertube upon which momentum states have been loaded thereby giving rise to three-charge black-hole microstates for which the corresponding black hole has a macroscopic horizon.  

From the gravitational perspective, such a closed null geodesic might lead to concerns  that there could  be CTC's.    However,  microstate geometries are stably causal and the closed null geodesic reflects the fact that a stationary observer at infinity (for asymptotically flat microstate geometries) becomes arbitrarily highly boosted when continued to the core of the microstate geometry \cite{Gibbons:2013tqa}.  It is this feature that leads to the long-term trapping as seen from infinity.
 
There is thus a lot  interesting physics associated with the neighborhood of the closed  null geodesic.  

There are also orbits in the heart of microstate geometries that intuition suggests should be loci for strong scrambling.  The microstate geometries known as superstrata can be characterized as a  capped BTZ geometry, $\cK$, with a deformed $3$-sphere, $S^3$, fibered over it.  For future reference, we use coordinates $(t, r, \psi)$ on $\cK$, where $\psi$ is the circle around the BTZ throat.  The cap is approximately ``the bowl'' at the bottom of a global AdS$_3$ and the $\psi$-circle pinches off smoothly at the  center of the cap.  The throat is where the circumference of the  $\psi$-circle stabilizes to a size determined by the momentum charge, $Q_P$, and the geometry is indeed very close to that of the BTZ solution.  There are, of course, deviations that appear as multipole moments in the BTZ throat and in the AdS-like cap. The one place where the geometry shows significant distortion is in the ``transition zone'' between the cap and the BTZ throat. This is where the momentum wave of the microstate geometry ``settles'' at some finite, but small value of $r$ determined by the angular momentum of the solution. Below the transition zone  (at lower values of $r$),  the geometry is like the AdS of the two-charge D1-D5 system while above the transition zone, the geometry feels the momentum charge, $Q_P$, and becomes the lower end of a BTZ throat.

Ultra-relativistic {\it infalling} probes already feel strong tidal forces from the multipoles in the BTZ throat, but the  tidal forces peak at the transition zone where the probe encounters the momentum wave sourcing the geometry.

The transition zone is also  very ``corrugated.''  Superstrata are sourced by momentum excitations, and these give rise to fields that oscillate around the $\psi$-circle of the BTZ throat and on the sphere, $S^3$.  In the Einstein frame,  and for solutions that are asymptotic to AdS$_3$,  the oscillations can be somewhat suppressed (coiffured) and even reduced to their RMS values for single modes, but they are strongly present in the string metric and for asymptotically flat solutions.   

One would expect that a string orbiting at the speed of light  around the $\psi$-circle, and on the $S^3$,  near the transition zone  would encounter these bumps as if they were a ``null shockwave.''   Indeed, this intuition is reinforced by the corresponding results arising from black holes and black rings.  In the early days of black ring construction, it was shown how one could seemingly put a varying charge density around the horizon of a black ring \cite{Bena:2004td}.  It was subsequently shown \cite{Horowitz:2004je} that the simplest such solutions are unphysical because the horizon could not be smooth: an infalling particle would spiral around such a varying charge density and experience it as a null shock wave.  One could soften the impact by coiffuring \cite{Bena:2014rea} but there was still a significant bump.  The correspondence between black holes and microstate geometries suggests that something similar should arise around the transition zone in the microstate geometry.  Certainly, infalling probes encounter huge tidal forces at the transition zone, but as we will see (at least for the superstratum considered in this paper),  strings orbiting at the speed of light avoid the very bumpy ride that one might have expected.  This happens because of some very simple physics.

We take the momentum wave sourcing the microstate geometry to  be left-moving. This biases the geometry.  We find that, for simple microstate geometries,  we can put a co-rotating closed geodesic ``orbit'' (or, more precisely, ``float'') at any fixed spatial position,  $(r,\psi)$, in the capped BTZ geometry.  However, the same is not true for  geodesics that are counter-rotating (relative to the background momentum wave).  For very small background momentum charge, $Q_P$, a counter-rotating geodesic can still be put anywhere, but there is a critical value of $Q_P$, above which the   counter-rotating geodesic cannot be placed above a finite radius, $r_{max}$:  the counter-rotating circular geodesics thus becomes bound to the geometry.  We show that this critical value of $Q_P$ is precisely  the boundary of the black-hole regime.  Thus, through its bound-state structure, the microstate geometry knows exactly where the black-hole bound lies.    

As $Q_P$ increases above the bound, the value of $r_{max}$ decreases rapidly so that, even for very modest values of $Q_P$, the  counter-rotating circular geodesics are confined to the AdS-like cap.  This means that, for any value of $Q_P$ in the black-hole regime, such geodesics have limited use as probes because they cannot explore the bumps associated with the transition zone.

On the other hand, the co-rotating geodesics can be placed anywhere.   However, these geodesics replicate the behavior of BPS particles and become truly co-moving, or floating,  with respect to the bumps in the space-time, $(t, r, \psi)$, directions.  Such particles thus sit still in the space-time corrugations rather than bounce around over them.  There is therefore no tidal enhancement associated with motion across the corrugations in the space-time.  On the other hand,  some of the co-rotating geodesics have non-trivial orbits on the sphere, $S^3$, and so can experience non-trivial bumps on this part of the geometry.

We will see that the scale of the tidal tensor, while fluctuating within the six-dimensional geometry, remains at a scale set by the AdS curvature.  The tidal stresses can still become large compared to the string scale, but this requires  the stringy probe to have a string-scale center of mass energy.   As we will discuss, such strings remain within the probe approximation and the corrugations in the metric means that strings develop a periodic, or oscillatory, mass term.  Such a mass term can create exponentially growing excitations of string modes and thus lead to the sort of chaotic scrambling one expects in a black hole. 

In Section \ref{Sect:MGs} we discuss the geometries and null geodesics of the superstrata that we are going to probe.   In Section  \ref{sec:Circles} we focus on  circular null geodesics, that is, null geodesics that make closed orbits on the deformed $S^3$ and run around the $\psi$-circle and fixed $r$ in the space-time, $\cK$. We are seeking to probe both the long-term trapping region of the geometry and the bumpiest part of the transition zone and  so we   fix the orbit on the $S^3$ at $\theta = \frac{\pi}{2}$, which is   the appropriate value for the supertube locus and  the value that maximizes  the bump functions.  From the perspective of the capped BTZ space-time, $\cK$, the angular momentum on the $S^3$ represents a Kaluza-Klein mass for the particle.  We classify the co-rotating and counter-rotating circular geodesics in this space-time, and show that the trapping of counter-rotating geodesics happens at  the black-hole bound. 

We discuss tidal forces in  Section  \ref{sec:Tides} and show how they remain small unless the probe energy approaches string scale.  We also show how the corrugations of the superstratum can result in oscillatory mass terms  for some of the string probes, and describe how tidal resonances can lead to exponentially growing string excitations with Lyapunov exponents that depend on the energy of the string, the amplitude of the mass term oscillation, and the proximity to a resonance.  This means that the scrambling will be chaotic.  Indeed, in a rather different context, it has already been shown how probes bouncing around the cores of microstate geometries can result in chaotic behavior \cite{Berenstein:2023vtd}.   Section \ref{sec:Conc} contains our final remarks.

\section{Probing Microstate geometries}
\label{Sect:MGs}

\subsection{The geometries and their null geodesics}
\label{ss:geoms}

 We are going to use the simplest work-horse of the microstate geometry program:  the $(1,0,n)$ superstratum \cite{Bena:2016ypk,Bena:2017upb,Bena:2017xbt,Heidmann:2019zws,Heidmann:2019xrd,Mayerson:2020tcl,Martinec:2020cml,Ceplak:2021kgl}. 
 
\subsubsection{The six-dimensional metric}
\label{ss:sixmet}

The six-dimensional part of the metric is most canonically written\footnote{There is a crucial typographical error in the coefficient of  $d\tau^2$ in $ \widehat{ds}_3^2$ in \cite{Martinec:2020cml} that is corrected in \cite{Ceplak:2021kgl}.  The computations in \cite{Martinec:2020cml} used the correct metric.} as a deformed $S^3$ fibration over a three-dimensional base manifold, $\cK$, that is asymptotic to AdS$_3$.  In the Einstein frame one has:
\begin{equation}
ds_6^2 ~=~ \sqrt{Q_1 Q_5} \, \Big(\, \widehat{ds}_3^2 ~+~ \widetilde{ds}_3^2 \,\Big)
\label{sixmet1}
\end{equation}
where
\begin{equation}
 \widehat{ds}_3^2  ~=~ \Lambda \, \bigg[ \frac{dr^2}{r^2 + a^2} ~+~\frac{r^2(r^2 + a^2)}{a^4}  \, d\psi^2 ~-~ \frac{1}{A^4  \,G} \, \bigg( d \tau +  \frac{A^2 r^2}{a^2} \, d \psi \bigg)^2\bigg]  \,, 
\label{Kmet1}
\end{equation}
and
\begin{equation}
\begin{aligned}
 \widetilde{ds}_3^2  ~=~ \Lambda \, d\theta^2 & ~+~  \frac{1}{\Lambda}\,  \sin^2 \theta \, \bigg(d \varphi_1  -  \frac{1}{ A^2}\, d \tau \bigg)^2 \\
 & ~+~  \frac{G}{\Lambda}\,  \cos^2 \theta\,  \bigg(d \varphi_2  +  \frac{1}{A^2 \, G}\,\Big( d\tau -\big (1+(A^2-1) F\big) \, d\psi \Big) \bigg)^2 \,.
 \end{aligned}
\label{Smet1}
\end{equation}
The functions $F$, $G$ and $\Lambda$ are ``bump functions:''
\begin{equation}
F ~\equiv~  1 - \frac{r^{2n}}{(r^2+a^2)^{n}} \,, \qquad   G ~\equiv~ 1 - \frac{a^2 \, b^2}{2 a^2+ b^2} \, \frac{ r^{2n}}{(r^2+a^2)^{n+1}}    \,, \qquad \Lambda ~\equiv~ \sqrt{ 1 - (1-G(r)) \, \sin^2 \theta  }  \,.
\label{FGLamdefn}
\end{equation}
The ``red-shift'' parameter, $A$, is defined by:
\begin{equation}
A ~\equiv~ \sqrt{ 1 +  \frac{b^2}{2 a^2}}  \,.
\label{Adefn}
\end{equation}

The $\psi$-coordinate is compactified on a unit circle: 
\begin{equation}
 \psi ~\equiv~ \psi ~+~ 2\pi \,, \label{psiperiod}
\end{equation}
and the time coordinate, $\tau$,  is dimensionless.   To make contact with the more standard formulation, one introduces a scale, $R_y$,  and the usual double null, $(u,v)$,  and space-time, $(t,y)$, coordinates:
\begin{equation}
  u ~=~  \coeff{1}{\sqrt{2}} (t-y)\,, \qquad v ~=~  \coeff{1}{\sqrt{2}}(t+y)~\equiv~  \frac{R_y}{\sqrt{2}} \, \psi \,, \qquad t ~=~ R_y \,\tau \,. \label{tyuv}
\end{equation}
The scale becomes  the radius of the $y$-circle:
\begin{equation}
  y ~\equiv~  y ~+~ 2\pi  R_y \,. \label{yperiod}
\end{equation}

Regularity of the solution requires
\begin{equation}
Q_1 Q_5 ~=~  \bigg(a^2 ~+~ \frac{b^2}{2} \bigg) \, R_y^2 \,.
\label{SSreg1}
\end{equation}
This leaves five remaining parameters:  the D1 and D5 brane charges, $Q_1, Q_5$,  two real parameters $a$ and $b$, and  the integer, $n \ge 0$, appearing in the bump functions.
The last three parameters  determine the angular momenta and the momentum charges of the solution:
\begin{equation}
\label{eq:ConsCharges}
 J_L  ~=~  J_R ~=~   \frac{R_y}{2} \,a^2\,,\qquad Q_P = \coeff{1}{2 } \, n \, b^2\,.
\end{equation} 

The supergravity solution is also supported by fluxes and one can find the precise expressions, and all the other relevant details, in earlier work, like \cite{Bena:2017upb,Bena:2017xbt,Heidmann:2019zws,Heidmann:2019xrd}.   However, to construct the string metric we will need the explicit forms of the electrostatic potentials:
\begin{equation}
\begin{aligned}
Z_1  ~=~ & \frac{Q_1}{\Sigma} \Big( 1 + (1-G(r))\,\sin^2 \theta\,    \cos (2\, n  \, \psi  +   2\, \varphi_1 ) \Big) \,,
\qquad Z_2  ~=~ \frac{Q_5}{\Sigma}  \,, \\    Z_4  ~=~ &     \frac{R_y}{\Sigma}\,\sqrt{(2a^2 + b^2) (1-G(r))} \, \sin  \theta \, \cos  ( n  \, \psi  +   \varphi_1 )  \,.
\end{aligned}
\label{Zforms}
\end{equation}
Indeed, one then has the canonical relationship with the warp factor, $\Lambda$: 
\begin{equation}
\Lambda ~\equiv~ \frac{\Sigma}{\sqrt{Q_1 Q_5}}\, \sqrt{Z_1 Z_2 ~-~ Z_4^2 }\,.
  \label{Lambdafrom}
\end{equation}

It is also very convenient to define a conformally related, six-dimensional metric by:
\begin{equation}
\begin{aligned}
{d\tilde s}_6^2  ~\equiv~  & \frac{1}{\sqrt{Q_1 Q_5} \, \Lambda}   \,ds^2_6 \\
 ~=~ &  \bigg( \frac{dr^2}{r^2 + a^2} ~+~\frac{r^2(r^2 + a^2)}{a^4}  \, d\psi^2 ~-~ \frac{1}{A^4 \,  G} \, \bigg( d \tau +  \frac{A^2 r^2}{a^2} \, d \psi \bigg)^2\bigg) \\
   &~+~d\theta^2 ~+~  \frac{1}{\Lambda^2}\,  \sin^2 \theta \, \bigg(d \varphi_1  -  \frac{1}{ A^2}\, d \tau \bigg)^2 \\
 & ~+~ \frac{G}{\Lambda^2}\,  \cos^2 \theta\,  \bigg(d \varphi_2  +  \frac{1}{A^2 \, G}\,\Big( d\tau -\big (1+(A^2-1) F\big) \, d\psi \Big) \bigg)^2  \,
\end{aligned}
\label{confsixmet}
\end{equation}
Indeed, we will work mainly with this metric.  Note that we have made it dimensionless by scaling out $\sqrt{Q_1 Q_5}$.  For $b=0$, this metric reduces to global AdS$_3$ with unit radius.

\subsubsection{Geodesic motion}
\label{ss:geodesics}

The  six-dimensional metric (\ref{sixmet1}) is independent of $(\tau,\psi,\varphi_1, \varphi_2)$, which means that the corresponding momenta are conserved:
\begin{equation}
L_1 ~=~ {K_{(1) M }} \frac{dz^M}{d \lambda} \,, \qquad L_2 ~=~ {K_{(2)  M }}  \frac{dz^M}{d \lambda} \,,\qquad  L_3 ~=~ {K_{(3)   M }}  \frac{dz^M}{d \lambda} \,, \qquad E ~=~ {K_{(4)   M }}  \frac{dz^M}{d \lambda}   \,,
  \label{ConsMom}
\end{equation}
where the $K_{(I)}$  are the Killing vectors: $K_{(1)}  = \frac{\partial}{\partial \varphi_1}$, $K_{(2)}  = \frac{\partial}{\partial \varphi_2}$, $K_{(3)}  = \frac{\partial}{\partial \psi }$ and $K_{(4)}  = \frac{\partial}{\partial \tau}$.   In addition, there is the standard quadratic conserved quantity coming from the metric: 
\begin{equation}
\varepsilon ~\equiv~ g_{MN} \, \frac{dz^M}{d \lambda} \, \frac{dz^N}{d \lambda}  \,.
  \label{MetInt}
\end{equation}

It was also shown in \cite{Bena:2017upb} that there is a conformal Killing tensor:
\begin{equation}
\Xi ~\equiv~ \xi_{MN} \, \frac{dz^M}{d \lambda} \, \frac{dz^N}{d \lambda} ~\equiv~Q_1 Q_5 \, \Lambda^2 \, \bigg( \frac{d\theta}{d \lambda} \bigg) ^2 ~+~ \frac{L_1^2}{\sin^2 \theta}~+~ \frac{L_2^2}{\cos^2 \theta} \,,
  \label{ConKtens}
\end{equation}
which, for any geodesic, satisfies: 
\begin{equation}
 \frac{d}{d \lambda}\Xi ~=~R_y\,\bigg( \frac{d\theta}{d \lambda} \bigg)  \, \bigg(\frac{\partial \Lambda}{\partial \theta}\bigg) \, \bigg( g_{MN} \, \frac{dz^M}{d \lambda} \, \frac{dz^N}{d \lambda}\bigg) \,.
  \label{ConKtens2}
\end{equation}

For null geodesics we therefore have six conserved quantities given by (\ref{ConsMom}), (\ref{MetInt})  and (\ref{ConKtens}).  Moreover,  null geodesics are independent of conformal transformations, and so we could equally well use the conformally related metric (\ref{confsixmet}).   Indeed, henceforth, and unless otherwise stated, we will work with (\ref{confsixmet}).  Again, note that this metric is scaled so that it is asymptotic to AdS$_3$ of {\it unit radius} at infinity.

In the metric (\ref{confsixmet}), and for null geodesics, the Killing tensor gives the conserved quantity:
\begin{equation}
\widetilde \Xi ~\equiv~\tilde  \xi_{MN} \, \frac{dz^M}{d \lambda} \, \frac{dz^N}{d \lambda} ~=~ \bigg( \frac{d\theta}{d \lambda} \bigg) ^2 ~+~ \frac{L_1^2}{\sin^2 \theta}~+~ \frac{L_2^2} {\cos^2 \theta}  ~=~ m^2 \,.
  \label{tConKtens}
\end{equation}
and metric conservation law can then be written  in the form:
\begin{equation}
\begin{aligned}
 & \frac{1}{r^2 + a^2} \bigg( \frac{dr}{d \lambda} \bigg)^2 ~+~\frac{a^4}{r^2  (r^2 + a^2)} \, \bigg( L_3 - \frac{r^2}{a^2} (L_1 + A^2 E)~+~    \frac{1}{A^2 G(r)}  \bigg( 1 +  \frac{b^2}{2 a^2}\,F(r) + A^2 \frac{r^2}{a^2}   \bigg) L_2 \bigg)^2    \\
&~-~\frac{1}{G(r)} \,\Big(L_2 - G(r)( L_1 + A^2 E)   \Big)^2 ~-~\big (1 - G(r)\big) \bigg( L_1^2 ~-~ \frac{ L_2^2 }{G(r)} \bigg)   ~=~ - m^2 \,.
\end{aligned}
  \label{AdS3motion}
\end{equation}
Recall that  affine parameters are not conformally invariant and one should therefore note that $\lambda$ is affine for the metric  (\ref{confsixmet}).

Remarkably, the dynamics of null geodesics are fully separated between the (deformed) sphere directions and  the (deformed) AdS$_3$ directions.   In particular, the  $\theta$ and $r$ dynamics are completely independent.  Moreover, the dynamics on the deformed sphere directions are identical to the those of the undeformed, round $S^3$.    We have introduced the parameter, $m$, as the energy of the motion on the sphere, and  from the perspective of the three-dimensional space-time, $\cK$, the $S^3$-energy, $m$, becomes the ``Kaluza-Klein'' mass of the particle.  As is evident from (\ref{AdS3motion}), the dynamics of these particles are complicated.

\subsubsection{The string metric}
\label{ss:stringmet}

Since we are going to consider string probes as well as geodesics, we will need the full ten-dimensional string metric.   This may be found in \cite[Appendix E.7]{Giusto:2013rxa} and  \cite{Bena:2017xbt}\footnote{There is a typographical error in  \cite{Bena:2015bea}: the warp factor in from of the six-dimensional metric should be $\sqrt{\alpha}$ and not $(\sqrt{\alpha})^{-1}$.}
\begin{equation}
d s^2_{10} ~=~\frac{ \sqrt{Z_1 Z_2}}{ \sqrt{Z_1 Z_2 - Z_4^2}}  \,ds^2_6 ~+~ \sqrt{\frac{Z_1}{Z_2}}\,d \hat{s}^2_{4}  ~=~ \Pi \, \Bigg(\sqrt{ Q_1 Q_5 }  \ {d\tilde s}_6^2 ~+~ \sqrt{\frac{Q_1}{Q_5}}\,d \hat{s}^2_{4} \Bigg) \,,
  \label{stringmet}
\end{equation}
where $d \hat{s}^2_{4}$ is the flat metric on $\IT^4$ and 
\begin{equation}
\Pi ~\equiv~ \sqrt{\frac{Q_5}{Q_1} \, \frac{Z_1}{Z_2}}  ~=~\sqrt{1 + (1-G(r))\,\sin^2 \theta\,    \cos (2\, n  \, \psi  +   2\, \varphi_1 )}\,.
  \label{PiDefn}
\end{equation}

Again, because null geodesics are conformally invariant, we can ignore the factors of $\Pi$ in  (\ref{stringmet}) and work with the much simpler dimensionless metric:
\begin{equation}
{d\tilde s}_6^2 ~+~ {\frac{1}{Q_5}}\,d \hat{s}^2_{4} \,.
  \label{simpmet}
\end{equation}

However, it will be important to keep track of affine parameters.  Suppose $\lambda$ is the affine parameter for   (\ref{simpmet}), and is the affine parameter inherent in
(\ref{tConKtens}) and  (\ref{AdS3motion}), and let $\aff$ be the affine parameter in the string metric (\ref{stringmet}), then we have:
\begin{equation}
\frac{d \aff}{d \lambda} ~=~ \Pi ~=~ \Big( 1 + (1-G(r))\,\sin^2 \theta\,    \cos (2\, n  \, \psi  +   2\, \varphi_1) \Big)^{\frac{1}{2}}\,.
  \label{affreln}
\end{equation}

It is also useful to recall how the curvature tensor behaves under conformal transformations of a metric.
Suppose one has 
\begin{equation}
d\hat s^2  ~\equiv~  \hat g_{\mu \nu} \, dx^\mu dx^\nu ~=~ \Omega^2 \, g_{\mu \nu} \, dx^\mu dx^\nu ~\equiv~  \Omega^2 \, ds^2\,,
  \label{conftrf}
\end{equation}
then the curvature, $\hat R^\rho{}_{\sigma\mu \nu}$ , of $\hat g_{\mu \nu}$ is given by:
\begin{equation}
\begin{aligned}
\hat R^\rho{}_{\sigma\mu \nu} ~=~    R^\rho{}_{\sigma\mu \nu}  ~-~\Big[ &  \delta^\rho_\mu \, \big( \nabla_\sigma V_\nu  - V_\sigma V_\nu \big) ~-~  \delta^\rho_ \nu \, \big( \nabla_\sigma V_\mu -   V_\sigma V_\mu \big)\\
 &~-~  g_{\sigma \mu} \,\big( \nabla^\rho V_\nu  -  V^\rho V_\nu\big) ~+~  g_{\sigma \nu} \,\big( \nabla^\rho V_\mu   -  V^\rho V_\mu \big)  \\
 &~+~ \big( V^\lambda V_\lambda \big)\,  \big(  \delta^\rho_\mu \,  g_{\sigma \nu} - \delta^\rho_\nu \, g_{\sigma \mu} \big) \Big]\,,
\end{aligned}
  \label{conftrfRiem}
\end{equation}
where $R^\rho{}_{\sigma\mu \nu}$ is the curvature of $g_{\mu \nu}$, all the index raising and lowering on the right-hand side is done in the metric $g_{\mu \nu}$, and where
\begin{equation}
\begin{aligned}
V_\mu ~\equiv~     \nabla_ \mu \, \log (\Omega) \,.
\end{aligned}
  \label{Vdefn}
\end{equation}
%

\subsection{The probes}
\label{ss:probes}

We are going to look at Penrose limits of the metric in the neighborhood of   ultra-relativistic particles.   We therefore start from  null geodesics, which, from  the discussion above,  may be thought of as a particular class of massive particles on $\cK$.   

\subsubsection{Null geodesic deviation and the Penrose limit}
\label{ss:NGD-PL}

To make contact with ultra-relativistic string probes, we will need the Brinkmann form of the Penrose limit, in which the ten-dimensional metric is reduced to the form of a plane-fronted gravitational wave:
\begin{equation}
ds^2 ~=~   2\, du \, dv ~+~ \Big( A_{ab}(u) x^a x^b \Big) \, du^2 ~+~\delta_{ab}\, d x^a d x^b     \,.
  \label{BPenrose1}
\end{equation}
The original construction of the Penrose limit was rather laborious.  One started with a null geodesic and then constructed a pencil of null geodesics in its neighborhood. One then scaled the solution so as to extract the second order expansion of the ultra-relativistic limit of the metric in the neighborhood of the original null geodesic, and then one had to make a non-trivial change of coordinate to get to the Brinkmann form of this metric. (For a review, see \cite{Blau:2002dy, Blau:2002mw}.)

Fortunately, this whole procedure has been greatly streamlined \cite{Blau:2003dz,Blau:2004yi,Blau:2006ar} in a manner that closely resembles geodesic deviation for time-like geodesics.  One starts from the original null geodesic, $x^M(u)$, where $u$ is an affine parameter.  Along this geodesic one constructs the Fermi transport (which is the same as parallel transport for geodesics parametrized by an affine parameter) of a set of orthonormal frames, $E^A \equiv {E^A}_M dx^M $:
\begin{equation}
\frac{d x^M}{du} \nabla_M \big(  {E^A}_P \big) ~=~  0 \,, \qquad g^{MN}  {E^A}_M {E^B}_N  ~=~  \eta^{AB}   \,,
  \label{PTframes}
\end{equation}
where $ \eta_{AB}$ is the Minkowski metric.  
Along the geodesic, $\gamma$, one may thus write:
\begin{equation}
ds^2 \big|_\gamma  ~=~  2 E^+ E^- ~+~ \delta_{ab}\,  {E^a}  {E^b}     \,.
  \label{metexp}
\end{equation}
One can then show that the matrix $A_{ab}$ in   (\ref{BPenrose1}) can be obtained from:
\begin{equation}
A_{ab} ~=~  - R_{MNPQ} \, {E^M}_a \, {E^P}_b\, \frac{d x^N}{du} \,\frac{d x^Q}{du}  ~\equiv~  - R_{a u b u}  \,.
  \label{Amat}
\end{equation}

As noted in \cite{Blau:2004yi}, this matrix governs the   transverse null geodesic deviation:
\begin{equation}
\frac{d^2}{du^2} \, Z^a ~=~ A_{ab}  \, Z^b  \,.
  \label{Deviation}
\end{equation}
where $Z^a$ is the transverse geodesic deviation vector.

The importance of the plane-fronted gravitational wave metric (\ref{BPenrose1}) is that string dynamics can be solved exactly in such backgrounds \cite{Horowitz:1990sr}.   Indeed, these techniques have been applied to microstate geometries and have led to remarkable new insights in terms of tidal trapping \cite{Martinec:2020cml,Ceplak:2021kgl}.
In this context, the significance of $A_{ab}$ is that it represents the negative of a mass matrix for the string.  Just as in the geodesic deviation equations, positive eigenvalues of $A_{ab}$ represent instabilities:  exponentially  growing deviations of geodesics, or negative masses for string excitations.  One way to drive instabilities in the bosonic string is if these negative masses  become large and overcome the string tension, as they do with infalling geodesics \cite{Martinec:2020cml,Ceplak:2021kgl}.   As we will discuss in Section \ref{ss:resonances}, a  periodic mass term can also drive unstable ``resonances'' in the string.  
 
\subsubsection{Conformal transformations of the null geodesic deviation}
\label{ss:conftrfNGD}

We consider two metrics, $\hat g_{\mu \nu}$ and $g_{\mu \nu}$, related by a conformal transformation as in   (\ref{conftrf}).   The starting point will be a null geodesic $x^\mu(\aff)$ with affine parameter, $\aff$, and  in the metric $\hat g_{\mu \nu}$ and affine parameter, $\lambda$, and  in $g_{\mu \nu}$.  One then has
\begin{equation}
\frac{d \aff}{d \lambda} ~=~ \Omega^2 \,.
  \label{affreln2}
\end{equation}
We introduce frames $\hat E^\pm, \hat E^a$  and $E^\pm,  E^a$ satisfying   (\ref{metexp}) for these two metrics along the null geodesic.   We take the inverse frames $\hat E_+$ and $ E_+$ to be tangent to the geodesic:  
\begin{equation}
\hat E^\mu_+  ~=~ \frac{d x^\mu}{d \aff} ~=~ \Omega^{-2}\, \frac{d x^\mu}{d \lambda} ~=~ ~ \Omega^{-2}\, E^\mu_+   \,.
  \label{plusinvframe}
\end{equation}
It follows that $\hat E^- = \Omega^{2}E^-$.  As stipulated in Section \ref{ss:NGD-PL}, we will take the frames $\hat E^A$ to be parallel transported along the null geodesic, however this is not a conformally invariant notion.   Indeed, if $\hat W^\mu$ is parallel transported in along the geodesic in $\hat g_{\mu \nu}$, this becomes
\begin{equation}
0  ~=~ \frac{d x^\rho}{d \aff} \hat \nabla_\rho \hat W^\mu  ~=~ \Omega^{-1} \frac{d x^\rho}{d \lambda}  \Big[\nabla_\rho W^\mu   - g^{\mu \sigma} (\nabla_\sigma \log\Omega)   W_\rho +  (\nabla_\sigma \log\Omega )W^\sigma  \delta^\mu_\rho  \Big]   \,,
  \label{parallelconf}
\end{equation}
where $W^\mu = \Omega \hat W^\mu$.  Parallel transport is only conformally covariant for vectors that satisfy: 
\begin{equation}
 W^\rho  \, \nabla_\rho \log\Omega ~=~  W_\rho  \,  \frac{d x^\rho}{d \lambda} ~=~  0    \,,  \qquad    W^\mu ~=~ \Omega \,\hat W^\mu \,.
  \label{orthogconds1}
\end{equation}
In particular, we note that this is true for the $\IT^4$ directions in   (\ref{stringmet})  and   (\ref{simpmet}).  Let $\hat E^\alpha = \Omega E^\alpha$ denote the parallel transported frames in these direction.

Using  (\ref{conftrfRiem}) and the orthogonality conditions in (\ref{orthogconds1}), it follows that along the  $\IT^4$ directions: 
\begin{equation}
\begin{aligned}
A^\alpha{}_\beta & ~=~  - \hat R^\alpha{}_{\mu \beta \nu} \, \frac{d x^\mu}{d \aff} \,\frac{d x^\nu}{d \aff} \\ &  ~=~  \delta^\alpha_\beta \  \Omega^{-4}\, \Big[ \nabla_\mu  \nabla_\nu  \log\Omega  - (\nabla_\mu  \log\Omega)(\nabla_\nu  \log\Omega) \Big] \, \frac{d x^\mu}{d \lambda} \,\frac{d x^\nu}{d \lambda}  
~=~ - \delta^\alpha_\beta \  \Omega^{-3} \, \frac{d^2}{d \lambda^2 } \bigg(\frac{1}{\Omega} \bigg)  \,. 
\end{aligned}
  \label{AmatT4}
\end{equation}
 To arrive at  this result  one must use the geodesic equation and  remember that the covariant derivatives in  (\ref{AmatT4})  are those of  (\ref{simpmet}) and thus the geodesic equation is most simply applied in  (\ref{simpmet}) using the corresponding affine parameter, $\lambda$.
 
 Making the change of affine parameter from $\aff$ to $\lambda$ on the left-hand side of  (\ref{Deviation}) yields the equation:
\begin{equation}
\frac{d^2}{d\aff^2} \, Z^\alpha ~=~ \Omega^{-4} \, \bigg[  \frac{d^2Z^\alpha }{d\lambda^2} ~-~ \frac{2}{\Omega}\, \bigg( \frac{d \Omega}{d\lambda}  \bigg) \, \frac{d Z^\alpha}{d  \lambda}   \bigg]  ~=~  -  \  \Omega^{-3} \,\bigg[ \frac{d^2}{d \lambda^2 } \bigg(\frac{1}{\Omega} \bigg)\bigg] \,  Z^\alpha \,.
  \label{DeviationT4a}
\end{equation}
Define  
\begin{equation}
 Z^\alpha ~=~  \Omega\, \tilde  Z^\alpha \,.
  \label{Ztdefn}
\end{equation}
and the geodesic deviation equation along the torus becomes:
\begin{equation}
 \frac{d^2 }{d\lambda^2} \tilde Z^\alpha~=~ 0  \,.
  \label{DeviationT4b}
\end{equation}
In other words, the deviation along the $\IT^4$ is entirely determined by the conformal rescaling.   The other directions are much more complicated and so we will ultimately resort to a more qualitative discussion.

\section{Circular orbits} 
\label{sec:Circles}

We now focus on the purely circular orbits in the $(1,0,n)$  microstate geometries.  For a round sphere there is obviously no distinction between any of the great circles, but for the superstratum, the most interesting locus lies at the peak of the bump functions and warp factors, and so we consider the circles at $\theta = \frac{\pi}{2}$.  This locus is also that of the closed null geodesic  of the underlying supertube.

\subsection{Simplifying the geodesic motion}
\label{ss:simpgeos}

From   (\ref{tConKtens}), it is evident that if we want to set $\theta = \frac{\pi}{2}$, we must take $L_2 = 0$.  The geodesics then only move in the $\varphi_1$ direction on the sphere and (\ref{tConKtens}) yields $L_1 = m$.  It is also natural to parametrize the $\psi$-motion in relation to $\varphi_1$ motion, or the Kaluza-Klein mass $m$.    Because of the mixing of the $\tau$ and  $\varphi_1$, the energy $E$ only appears in the combination $L_1 + A^2 E$.  We will assume $m \ne 0$, and  take: 
\begin{equation}
L_1  ~=~ m \,, \qquad  L_2  ~=~ 0 \,, \qquad L_3  ~=~ - \gamma \, m  \,, \qquad \hat E  ~=~1+  \frac{A^2 E}{m} \,,
  \label{reparam}
\end{equation}
for some choice of $L_1$, $\hat E$ and $\gamma$.

The radial motion can now be characterized through an effective potential:
\begin{equation}
\bigg( \frac{dr}{d \lambda} \bigg)^2 ~+~ m^2 \, V(r; \hat E, \gamma)    ~=~ 0 \,,  
  \label{radialmotion}
\end{equation}
where
\begin{equation}
 V(r; \hat E, \gamma)    ~=~  \frac{1}{r^2} \big(\hat E \, r^2 +  \gamma \, a^2 \big)^2 ~+~(r^2 + a^2)\, G(r)\,\big(1 - \hat E^2 \big)   \,.
   \label{potential}
\end{equation}
From (\ref{radialmotion}) it is evident that we can absorb $m$ into a rescaling of the affine parameter, $\lambda$.  However, as one can see from (\ref{reparam}), the sign of $m$ plays a role in the form of the geodesic motion and so we will retain $m$, allowing $m = \pm 1$.

For circular geodesics one must  have  $V(r; \hat E, \gamma) = 0$ and $\frac{d}{dr} V(r; \hat E, \gamma) = 0$, which give the following two constraints:
\begin{align}
& \frac{1}{r^2} \,\big(\hat E \, r^2 +  \gamma \, a^2 \big)^2 ~+~(r^2 + a^2)\, G(r)\,\big(1 - \hat E^2 \big) ~=~  0  \,,
   \label{constr1} \\ 
&  \frac{2}{r^3} \,\big(\hat E^2 \, r^4 -  \gamma^2 \, a^4 \big)    ~+~  (1 -  \hat E^2) \, \big(2\, r \, G(r) +  (r^2 + a^2)\, G'(r)  \big) ~=~ 0  \,.
   \label{constr2}
\end{align}
These two equations determine $\hat E$ and $\gamma$ in terms of the constant value of $r$  for circular orbits.   By taking linear combinations of  (\ref{constr1}) and (\ref{constr2}), one can obtain a linear equation for $\gamma$. If one then substitutes this back into  either (\ref{constr1}) or (\ref{constr2}), one obtains a quadratic equation for $\hat E^2$:
\begin{equation}
\big(\hat E^2 -1\big)\,\Big[ \big(P(r)^2  ~-~ 16\, r^2 (r^2 + a^2) \,G(r) \big) \, \hat E^2     ~-~ P(r)^2   \Big]  ~=~   0   \,,  
   \label{quadratic}
\end{equation}
where
\begin{equation}
P(r)  ~\equiv~ r (r^2 + a^2) \,G'(r) + 2 \, (2\,r^2 + a^2)\,G(r) \,,
   \label{Pdefn}
\end{equation}
and $\gamma$ is given by 
\begin{equation}
\gamma ~=~ - \frac{1}{ 4\, a^2\, \hat E  } \Big[ \, 4\, r^2\, \hat E^2  ~-~  \big(\hat E^2 -1\big) \, P(r)\, \Big] \,.
   \label{gammasub}
\end{equation}

The velocities are then given by: 
\begin{equation}
\begin{aligned}
\frac{d r}{d \lambda}  ~\equiv~ & 0 \,, \qquad  \frac{d \tau}{d \lambda}~=~ m \, A^2 \, \bigg[ \, \frac{(\hat E\, r^2 + \gamma \, a^2)}{(r^2 + a^2)} ~-~ \hat E\, G(r)   \,  \bigg]  \,, \qquad \frac{d \psi}{d \lambda} ~=~  - m \,   \frac{a^2\,( \hat E\, r^2 + \gamma \, a^2)}{r^2(r^2 + a^2)}  \,, 
\\
 \frac{d \theta}{d \lambda}  ~\equiv~ & 0  \,, \qquad    \frac{d \varphi_1}{d \lambda} ~=~    m \,   \bigg[ \, \frac{(\hat E\, r^2 + \gamma \, a^2)}{(r^2 + a^2)} ~+~ (1-\hat E)\, G(r)   \,  \bigg]\,, \qquad   \frac{d \varphi_2}{d \lambda} ~=~   - m \,   \frac{ \gamma \, a^2}{r^2}      \,.
\end{aligned}
   \label{geovels}
\end{equation}
Given the velocity combinations that appear in (\ref{Smet1}), it is worth noting that the boosted angular velocity around the $S^3$ follows the profile function, $G(r)$:
\begin{equation}
\frac{d \varphi_1'}{d \lambda}  ~\equiv~   \frac{d \varphi_1}{d \lambda} ~-~ \frac{1}{A^2} \, \frac{d \tau}{d \lambda}~=~  m \,G(r)   \,.
   \label{phi1tauvel}
\end{equation}

Combining the fact that there are four roots for $\hat E$, and the fact that we can absorb the magnitude of $m$ into $\lambda$, so that $ m = \pm 1$, there are, in principle, eight choices for circular geodesics.  
The metric and geodesic equations have an obvious symmetry under $\tau \to - \tau, \psi \to - \psi$, and we want to restrict to future-directed geodesics ($d\tau >0$), thus there are only four physically interesting distinct solutions.  The choices are related to the orientations of the rotations in the $\psi$ and $\varphi_1$ directions.  This is easiest to illustrate by going back to the pure AdS$_3$  solution.

\subsection{The AdS$_3$ limit}
\label{ss:AdS3}

The pure AdS$_3$ solution is obtained by setting $b =0$, and hence $G(r) \equiv 1$.  It is useful to note that in this limit, the metric (\ref{confsixmet}) becomes:
\begin{equation}
\begin{aligned}
{d\tilde s}_6^2  
 ~=~ &   \frac{dr^2}{r^2 + a^2} ~+~\frac{r^2}{a^2}  \, (d\psi- d\tau )^2 ~-~\frac{r^2+ a^2}{a^2}  d \tau^2  \\
 & ~+~ d\theta^2 ~+~  \sin^2 \theta \, (d \varphi_1  -  d \tau )^2 
  ~+~    \cos^2 \theta\,  \big(d \varphi_2   -   (  d\psi - d\tau ) \big)^2  \\
  ~=~ &   \frac{dr^2}{r^2 + a^2} ~+~\frac{r^2}{a^2}  \,  d\psi'{}^2 ~-~\frac{r^2+ a^2}{a^2}  d \tau^2  ~+~ d\theta^2 ~+~  \sin^2 \theta \,  d \varphi_1'{}^2  \,,
  ~+~    \cos^2 \theta\,   d \varphi_2'{}^2  \,
\end{aligned}
\label{AdSmet}
\end{equation}
where $\psi'  =\psi - \tau$,  $\varphi_1'  =\varphi_1 - \tau$ and $\varphi_2'  =\varphi_2 -(\psi - \tau)$. 

The four solutions to the constraints are then:
\begin{equation}
( \hat E\, \,, \, \gamma)~=~   \pm \Big( 1 \, \,, \, - \frac{r^2}{a^2} \Big)  \,, \qquad ( \hat E\, \,, \, \gamma)    ~=~   \pm \Big( 1+ \frac{2\,r^2}{a^2} \, \,,  \, \frac{r^2}{a^2} \Big)  \,,
   \label{AdSsols1}
\end{equation}
and the corresponding velocities are:
\begin{equation}
\begin{aligned}
\bigg( \frac{d \tau}{d \lambda}, \frac{d\psi}{d \lambda}, & \frac{d\varphi_1}{d \lambda}, \frac{d\varphi_2}{d \lambda} \bigg) \\
&  ~=~   - m \,  \big(1 \,,0 \,, 0\,,-1\big ) \,, \  \ m \,  \big(1 \,,0 \,, 2\,,-1\big ) \,, \ \   -m \,  \big(1 \,, 2 \,, 0\,, 1\big ) \,,   \ \  m \,  \big(1 \,,2 \,, 2\,, 1\big ) \,, 
\end{aligned}
   \label{AdSsols2}
\end{equation}
which may be written as:
\begin{equation}
\begin{aligned}
\bigg( \frac{d \tau}{d \lambda}, \frac{d\psi'}{d \lambda}, & \frac{d\varphi_1'}{d \lambda}, \frac{d\varphi_2'}{d \lambda} \bigg) \\
&  ~=~   - m \,  \big(1 \,,-1 \,, -1\,, 0 \big ) \,, \ \ m \,  \big(1 \,,-1 \,, 1\,, 0 \big ) \,, \   \  -m \,  \big(1 \,, 1 \,, -1\,,  0 \big )   \,, \ \ m \,  \big(1 \,,1\,, 1\,, 0\big ) \,.
\end{aligned}
   \label{AdSsols3}
\end{equation}

We will always choose the sign of $m$ so that $\frac{d \tau}{d \lambda} >0$.  The four solutions then correspond to the four choices of the direction  of rotations in the  $\psi'$ and $\varphi_1'$ directions.  The values of $\frac{d\varphi_2}{d \lambda}$ in   (\ref{AdSsols2}) are simply an artifact of setting $L_2 =0$ and the mixing terms in   (\ref{AdSmet}).

\subsection{Circular orbits in the superstratum}
\label{ss:COss}

\subsubsection{The ``BPS geodesics'' }
\label{ss:BPSgeodesics}

One set of superstratum orbits is extremely simple, and closely resemble the co-rotating AdS$_3$ geodesics:
\begin{equation}
( \hat E\, \,, \, \gamma)~=~   \pm \Big( 1 \, \,, \, - \frac{r^2}{a^2} \Big) \,.
   \label{simpsols1}
\end{equation}
The corresponding velocities are:
\begin{equation}
\bigg( \frac{d \tau}{d \lambda}, \frac{d\psi}{d \lambda},   \frac{d\varphi_1}{d \lambda}, \frac{d\varphi_2}{d \lambda} \bigg) 
~=~   - m \,  \big(A^2 \, G(r) \,,0 \,, 0\,,-1\big ) \,, \qquad   m \,  \big(A^2 \, G(r)  \,,0 \,, 2\, G(r) \,, -1\big )  \,.
\label{ssvels1}
\end{equation}
Note that both these sets of geodesics have $ \frac{d\psi}{d \lambda} \equiv 0$, and so, from  the perspective of the three-dimensional space-time, $\cK$, these geodesics  are also co-rotating.  There is also no restriction on the value of $r$: these geodesics are smooth and well-defined for all values of $r$.  They are thus like BPS objects in that they can be placed anywhere within the superstratum.  Indeed, from the perspective of the manifold, $\cK$, they are floating branes \cite{Bena:2009fi} in that they sit happily at fixed $(r, \psi)$, feeling no force.  We will therefore refer to these orbits as ``BPS geodesics.''   We note that the first family of these geodesics has $\frac{d\varphi_1}{d \lambda} =0$ and is therefore also co-rotating, or floating, on the $S^3$.

As we will see, the counter-rotating geodesics have very different properties  and are ``very non-BPS'' objects.

\subsubsection{Bounding the counter-rotating geodesics }
\label{ss:countergeodesics}

The counter-rotating geodesics come from the values of $\hat E$ that lead to the vanishing of the expression in the square bracket of (\ref{quadratic}).  To have non-trivial solutions one must have 
\begin{equation}
Q(r) ~\equiv~ \big(P(r)^2  ~-~ 16\, r^2 (r^2 + a^2) \,G(r) \big)   ~>~   0   \,,  
   \label{coeff}
\end{equation}
This is true for $b=0$ and for small values of $b$.  However, as we will see, there is a critical value, $b_{crit}$, of $b$, above which $Q(r)$ can vanish, and become negative.  Indeed, we will show that if $b > b_{crit}$ then counter-rotating circular geodesics can only exist at values of $r$ less than a maximal value, $r_{max}$, that depends on $b$.

This result follows from the following identity

\begin{equation}
\begin{aligned}
Q(r)  ~=~    4\,a^4\, &  \bigg( 1 +  \frac{b}{\sqrt {2 a^2 + b^2}} \, x^n \, \Big(\sqrt {n+1 } ~+~  \sqrt {n } \, x \Big ) \bigg) \,  \bigg( 1 +  \frac{b}{\sqrt {2 a^2 + b^2}} \, x^n \, \Big(\sqrt {n+1 } ~-~  \sqrt {n } \, x \Big ) \bigg) \\
  \times & \bigg( 1 -  \frac{b}{\sqrt {2 a^2 + b^2}} \, x^n \, \Big(\sqrt {n+1 } ~+~  \sqrt {n } \, x \Big ) \bigg) \,  \bigg( 1 -  \frac{b}{\sqrt {2 a^2 + b^2}} \, x^n \, \Big(\sqrt {n+1 } ~-~  \sqrt {n } \, x \Big ) \bigg)\,, 
 \end{aligned}
   \label{Qidentity}
\end{equation}
where
\begin{equation}
x ~\equiv~  \frac{r}{\sqrt {r^2 + a^2}}   \,.
   \label{xdefn}
\end{equation}
Since $0 \le x < 1$, it follows that all the terms in the parentheses in   (\ref{Qidentity})  are strictly positive, except, possibly, for the third parenthesis:
\begin{equation}
q(r)  ~\equiv~  \bigg( 1 -  \frac{b}{\sqrt {2 a^2 + b^2}} \, x^n \, \Big(\sqrt {n+1 } ~+~  \sqrt {n } \, x \Big ) \bigg) \,.
   \label{qdefn}
\end{equation}
Define  $b_{crit}$ by the vanishing of $q(r)$ as $r \to \infty$.  One then finds that 
\begin{equation}
b_{crit}  ~=~  a \, \sqrt{\sqrt{1 + \frac{1}{n}} -1} \,.
   \label{bcritdefn}
\end{equation}
For $b > b_{crit}$, $q(r)$ will go negative for $r$  sufficiently large.  

Define $r_{max}$ by:
\begin{equation}
q(r_{max})  ~=~ 0  \,,
   \label{rmaxdefn}
\end{equation}
which has no solutions for $b < b_{crit}$ and a single real solution for  $b >  b_{crit}$.    It follows that counter-rotating circular geodesics can be located at any value of $r$ when $b \le b_{crit}$, while for $b > b_{crit}$ they can only be located at $r < r_{max}$. 

It is also very interesting to note that (i)  $b_{crit}$ is very small, and (ii) the value of  $r_{max}$ drops very rapidly from infinity to the cap as $b$ increases above $b_{crit}$.    To be more specific,  $b_{crit}$ decreases monotonically with $n$, starting at $a \sqrt{\sqrt{2} -1} \approx 0.6436 \, a$ and, at large $n$, it is asymptotic to $ \frac{a}{\sqrt{2 n}}$.  The transition zone between the cap and the BTZ throat is approximately at the turning point of  $G(r)$, or at $r = a \sqrt{n}$.  For $r_{max}$ to drop to this value of $r$ one must have $q(a \sqrt{n}) =0 $, which means %
\begin{equation}
 b ~=~ b_{cap} ~\equiv~a \, \sqrt{\frac{2 \, (n+1)^{n+1} }{(2n+1)^2 \,n^n - (n+1)^{n+1} }}  \,.
   \label{btrans}
\end{equation}
Thus the counter-rotating orbits are restricted to the cap for $b>b_{cap}$.  The expression for $b_{cap}$ is also a monotonically decreasing function of $n$, starting at $2 a \sqrt{\frac{2}{5}} \approx 1.2649 \, a$ and, at large $n$, it is asymptotic to $a\,  \sqrt{\frac{e}{2 n}}\approx   1.1658 \frac{a}{\sqrt{n}}$.

\begin{figure}[t]
\centering
\includegraphics[width=0.45\textwidth]{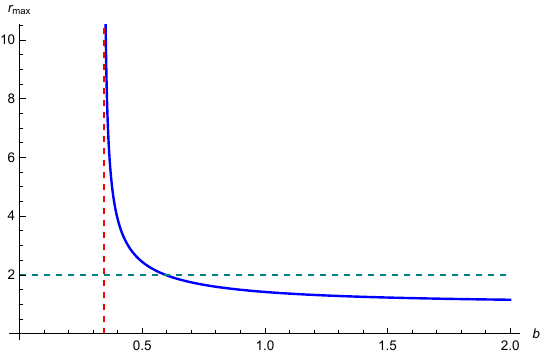} \qquad
\includegraphics[width=0.45\textwidth]{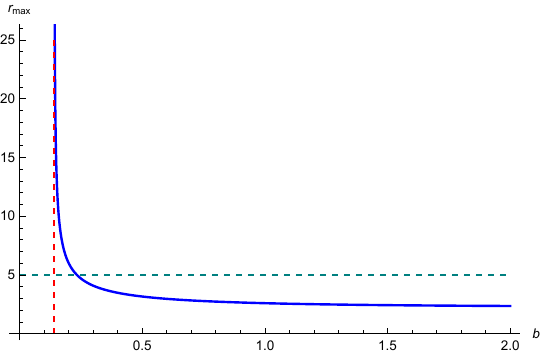} 
\caption{Plots of $r_{max}$ as a function of $b$ for counter-rotating circular geodesics with $a=1$ with $n=4$ (on the left) and $n=25$ (on the right). The  horizontal dashed line marks the transition  at $r = \sqrt{n} \, a$ between the BTZ throat and the cap.  The  vertical dashed line is at $b= b_{crit}$.  Up until $b= b_{crit}$ there is no limit on the radius of the counter-rotating geodesics, but as $b$ increases above $b_{crit}$, the maximum possible radius for circular counter-rotating geodesics, $r_{max}$, drops precipitously towards the end of the BTZ throat and the start of the cap. When scaled appropriately, the features of both curves are almost identical.}
\label{fig:rmax}
\end{figure}

 We have shown some illustrative examples of the behavior of $r_{max}$  as  a function of $b$ in Fig.~\ref{fig:rmax}.

The rapid confinement of counter-rotating geodesics to the cap of the microstate geometry suggests a critical transition to a non-BPS bound state. We now show that there is indeed a clear physical interpretation of the behavior we have found here.

\subsubsection{Holography and the bound on geodesics}
\label{ss:bounds}

It is valuable to recall the holographic dictionary for  the superstrata that we are using here \cite{Bena:2016ypk, Bena:2017xbt}.  These solutions  are dual to the D1-D5 state:
 \begin{equation}
(|\!+\!+\rangle_1)^{N_1 }
\Big((L_{-1}- J^3_{-1})^n|00\rangle_{k=1}\Big)^{N_{2} }\,,
  \label{CFTcstate}
\end{equation}
with the constraint  $N_1+  N_2 = N \equiv n_1 n_5$, where $n_1$ and $n_5$ are the numbers of underlying D1 and D5 branes.   The quantized angular momenta, $j_L, j_R$,  and momentum, $n_P$,  of this state are given by:
\begin{equation}
  \label{eq:jljr}
  j_L  ~=~  j_R  ~=~  \coeff{1}{2}\, N_1      \,, \qquad    n_P   ~=~  n\, N_2     \,.
\end{equation}
These are related to the supergravity parameters by: 
\begin{equation}
  j_L  ~=~  j_R   ~=~    \coeff{1}{2}\, \mathcal{N} \, a^2  \,, \qquad  n_ P ~=~  \coeff{1}{2}\, \mathcal{N} \, n \, b^2\,, 
 \label{jljrnp}
\end{equation}
where $ \mathcal{N} \equiv n_1 n_5 R_y^2 / (Q_1 Q_5)$.  The constraint $N_1+  N_2  = n_1 n_5$ then translates directly into the regularity condition (\ref{SSreg1}).   For these states one can also write the regularity condition as: 
\begin{equation}
 j    ~+~ \frac{n_P}{2\, n}   ~=~  \coeff{1}{2} \, n_1 n_5 \,. 
   \label{jnpcond}
\end{equation}
where $j \equiv  j_L = j_R$.

As noted in  \cite{Bena:2016ypk}, rotating D1-D5-P black holes with regular horizons exist when $n_1 n_5 \np-j^2 >0$ and this cosmic censorship bound  defines the ``black-hole regime'' for these parameters. This bound translates into  the supergravity parameters as:
\begin{equation}
 0 ~<~ n_1 n_5 n_P  - j^2  ~=~    \frac{1}{2}\,  \mathcal{N}^2 \bigg( \frac{Q_1 Q_5}{R_y^2} \, n \, b^2 \, - \frac{1}{2}\,  a^4 \bigg) ~=~\frac{1}{4}\,  \mathcal{N}^2 \, n \,\bigg( \,  \bigg(\frac{b^2}{a^2}~+~  1 \bigg)^2 ~-~ \bigg(1+ \frac{1}{n} \bigg)\bigg)     \,, 
\label{bhbound1}
\end{equation}
where we have used the regularity condition  (\ref{SSreg1}).  It follows that the black-hole bound is simply:
\begin{equation}
   \frac{b^2}{a^2} ~>~ \sqrt{1+ \frac{1}{n}} ~-~1 ~\equiv~   \frac{b_{crit}^2}{a^2}   \,,
\label{bhbound2}
\end{equation}
where $b_{crit}$ was introduced in   (\ref{bcritdefn}).

Thus the critical value of $b$ that determines the onset of trapping of counter-rotating circular geodesics is nothing other than the black-hole bound.

This is particularly interesting because, as $b$ increases, the overall superstratum geometry  slowly changes from global AdS$_3$, at $b=0$, to some capped BTZ geometry with a progressively longer  throat.  There is no sudden and obvious transition in the geometry as one crosses the black-hole bound.  However, we see that the counter-rotating geodesics know exactly where this bound lies and provide an ``order parameter'' for this transition.

We  conclude this section by looking at some representative examples of the geodesics.

\subsubsection{Orbits in the cap, at infinity and in between}
\label{ss:sols}

For $n>1$, the metric in the cap\footnote{For $n=1$, the bump functions do not quite die off fast enough as $r \to 0$ and so there is some deformation of the cap metric.} is  that of AdS$_3$.  If one drops all the $r^{2n}$ terms in  (\ref{confsixmet}) one obtains:
\begin{equation}
\begin{aligned}
{d\tilde s}_6^2  
 ~=~ &  \frac{dr^2}{r^2 + a^2} ~+~\frac{r^2}{a^2}  \,  \bigg( d\psi - \frac{d\tau }{A^2} \bigg)^2 ~-~ \frac{ (r^2 + a^2)}{a^2}  \bigg( \frac{d\tau }{A^2}   \bigg)^2 \\
   &~+~d\theta^2 ~+~   \sin^2 \theta \, \bigg(d \varphi_1  -  \frac{d \tau}{ A^2}  \bigg)^2  ~+~  \cos^2 \theta\,  \bigg(d \varphi_2  +\,\Big(  \frac{d\tau }{A^2}  -    \, d\psi \Big) \bigg)^2 \\
    ~=~ &  \frac{dr^2}{r^2 + a^2} ~+~\frac{r^2}{a^2}  \, d\psi'{}^2   ~-~ \frac{ (r^2 + a^2)}{a^2} \,d\tau'{}^2  ~+~d\theta^2 ~+~   \sin^2 \theta \,  d \varphi_1'{}^2    ~+~  \cos^2 \theta\,   d \varphi_2'{}^2    \,
\end{aligned}
\label{capmet}
\end{equation}
where $\tau' = \frac{\tau}{A^2}$,  $\psi'  =\psi - \tau'$,  $\varphi_1'  =\varphi_1 - \tau'$ and $\varphi_2'  =\varphi_2 -(\psi - \tau')$.  The only difference between this and the $b \to 0$ limit is the factor of $A^2$ rescaling $\tau$, which represents the redshift between the top and bottom of the superstratum throat. 

The  four solutions to the constraints are then exactly as in   (\ref{AdSsols1}) and these lead to orbital motion given by   (\ref{AdSsols2}) and  (\ref{AdSsols3}) except that $ \frac{d \tau}{d \lambda}$ must be replaced by  $\frac{d \tau'}{d \lambda}$.

At infinity, the metric is asymptotically AdS$_3$ and $G(r) \to 1$ as $r \to \infty$.  The  co-rotating circular geodesics with velocities (\ref{ssvels1}) simply limit to those of AdS$_3$, (\ref{AdSsols2}), up to the re-scaling  $\tau' = \frac{\tau}{A^2}$.  

The  story is somewhat different for the counter-rotating geodesics.   Expanding the solutions to the constraints around infinity  gives:
\begin{equation}
 ( \hat E\, \,, \, \gamma)    ~=~   \pm  \frac{r^2}{\sqrt{n} \, \sqrt{(b_{crit}^2 - b^2)(b_{crit}^2 + b^2 +2)}} \,  \big(\, 2\, A^2 \, \,,  \,  -1 \, \big)  \,,
   \label{infsols1}
\end{equation}
and the corresponding velocities are: 
\begin{equation}
\bigg( \frac{d \tau}{d \lambda}, \frac{d\psi}{d \lambda},  \frac{d\varphi_1}{d \lambda}, \frac{d\varphi_2}{d \lambda} \bigg)  
  ~=~   m\,  \big(0 \,,0 \,, 1\,,0\big )~\mp~ \frac{m\,  a^2}  { \sqrt{n} \, \sqrt{(b_{crit}^2 - b^2)(b_{crit}^2 + b^2 +2)}}   \,  \big( \,  A^2 \, \,,  \,2 \,  A^2 \, \,,  \,1  \, \,,  \,1 \, \big )   \,.
   \label{infsols2b}
\end{equation}
These match the AdS$_3$ results for counter-rotating geodesics,  (\ref{AdSsols1}) and  (\ref{AdSsols2}), for $b=0$.  These solutions also become unphysical for $b > b_{crit}$.

\begin{figure}[t]
\centering
\includegraphics[width=0.45\textwidth]{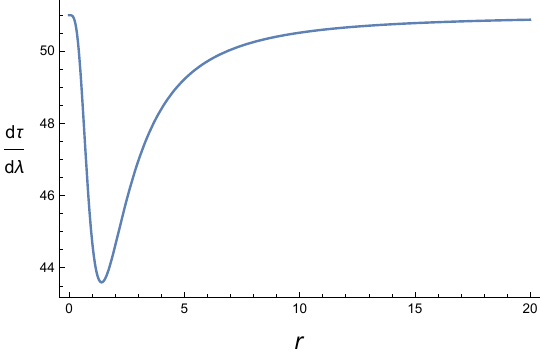} \qquad
\includegraphics[width=0.45\textwidth]{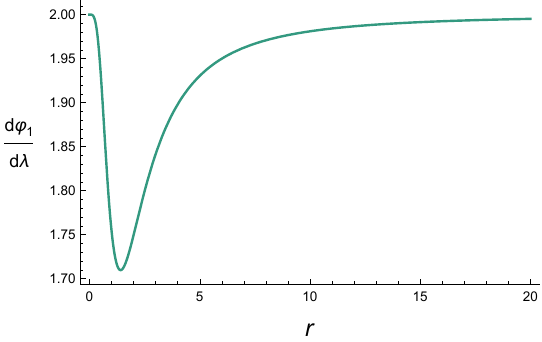} 
\caption{The geodesic velocities, $\big( \frac{d \tau}{d \lambda},   \frac{d\varphi_1}{d \lambda} \big)$, for the second set of co-rotating geodesics in (\ref{ssvels1}), with  $a=1$, $b=10$, $n=2$.  (We do not plot  $\big( \frac{d \psi}{d \lambda},   \frac{d\varphi_2}{d \lambda} \big)$ because these velocities are constant.) Note that the velocities are finite and well-defined for  $0 \le r < \infty$.  These velocities follow the profile of   $G(r)$, whose minimum is at $r = \sqrt{2} \approx 1.4142$. }
\label{fig:co-geodesics1}
\end{figure}

\begin{figure}[t]
\centering
\includegraphics[width=0.45\textwidth]{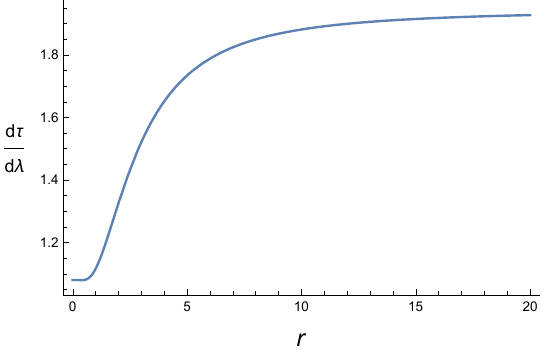} \qquad
\includegraphics[width=0.45\textwidth]{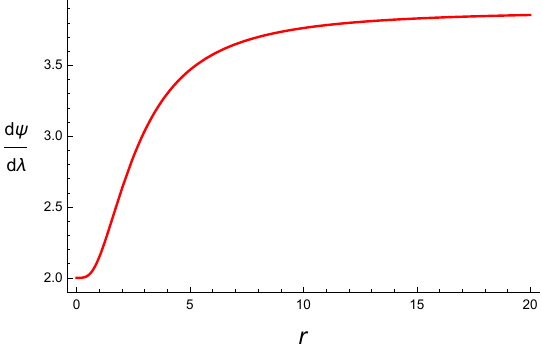} \\
\includegraphics[width=0.45\textwidth]{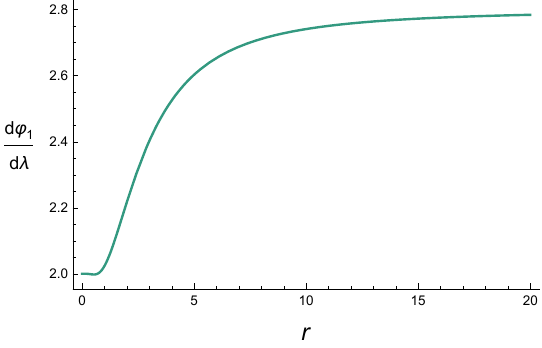}\qquad
\includegraphics[width=0.45\textwidth]{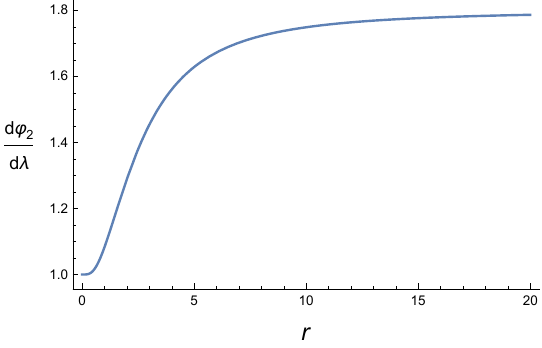}
\caption{Geodesic velocities for a counter-rotating geodesic with  $a=1$, $b=\frac{2}{5}$, $n=2$.  These geodesics exist for all values of $r$ because  $b < b_{crit} = \Big(\sqrt{\frac{3}{2}} -1\Big)^{\frac{1}{2}} \approx 0.4741$.}
\label{fig:counter-geodesics1}
\end{figure}

Finally, we  catalog some representative examples of the velocities of particles on circular geodesics of various radii. Fig.~\ref{fig:co-geodesics1} shows the non-trivial velocities of simple, but generic, ``co-rotating'' geodesic.  These geodesics  exist for all values of $r$.  Fig.~\ref{fig:counter-geodesics1} shows  a generic  ``counter-rotating'' geodesic for a value of $b < b_{crit}$.   Again these   exist for all values of $r$. 

\begin{figure}[t]
\centering
\includegraphics[width=0.45\textwidth]{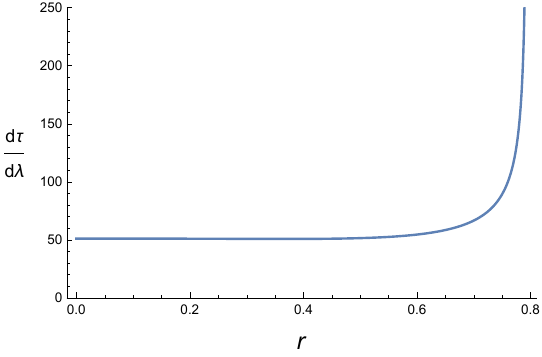} \qquad
\includegraphics[width=0.45\textwidth]{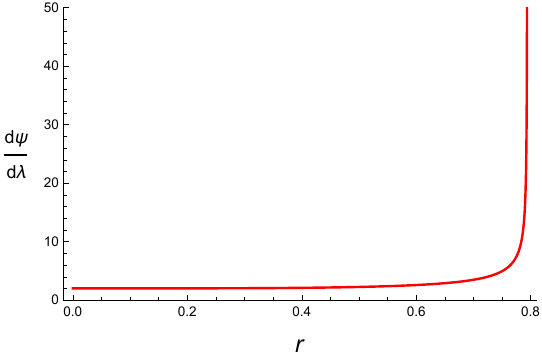} \\
\includegraphics[width=0.45\textwidth]{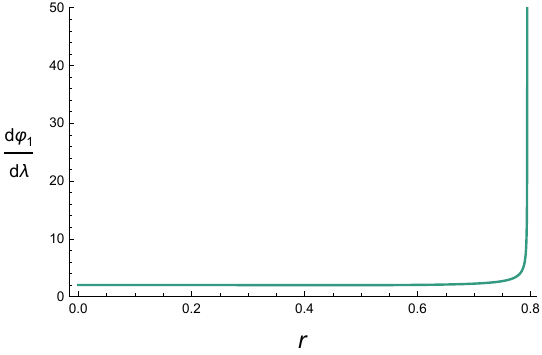}\qquad
\includegraphics[width=0.45\textwidth]{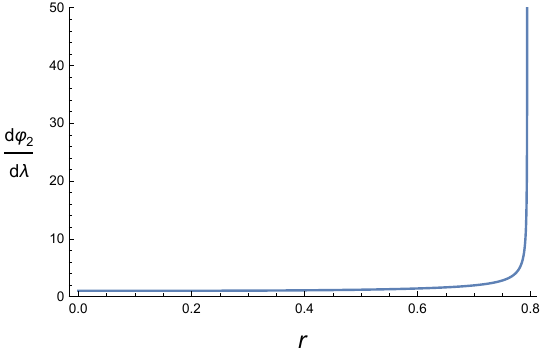}
\caption{Geodesic velocities for a counter-rotating geodesic with  $a=1$, $b=10$, $n=2$.   Here $b_{crit} = \Big(\sqrt{\frac{3}{2}} -1\Big)^{\frac{1}{2}} \approx 0.4741$ and for $b=10$ one has $r_{max} \approx 0.7941 $.  The geodesic velocities diverge as the orbit is pushed out to $r_{max}$. }
\label{fig:counter-geodesics2}
\end{figure}

Fig.~\ref{fig:counter-geodesics2} shows   a  generic ``counter-rotating'' geodesic for a value of $b >  b_{crit}$.   These geodesics can only exist for a limited range of values of $r < r_{max}$, and the velocities all diverge as $r \to r_{max}$.  For the example depicted, we have $b \gg  b_{crit}$, and circular  counter-rotating  geodesics can only exist in the cap region.

\section{Geodesic deviation, tidal forces and resonances }
\label{sec:Tides}

\subsection{The scale of the tidal forces}
\label{ss:scale}

To bound the size of the tidal forces, we define
\begin{equation}
M_{MN} ~=~  - R_{MNPQ} \frac{d x^N}{du} \,\frac{d x^Q}{du}  ~\equiv~  - R_{M u N u}  \,,
  \label{Mmat}
\end{equation}
and note that 
\begin{equation}
M_{MN}  \, \frac{d x^N}{du} ~=~ 0  \,.
  \label{Mnullvec}
\end{equation}
It follows that 
\begin{equation}
{\cal A}~\equiv~   \sqrt{A_{ab} \, A_{ab}} ~=~ \sqrt{M_{MN} M^{MN}}  \,.
  \label{TFmag}
\end{equation}
The latter expression is simpler to compute because one does not need the parallel-transported frames.  

Since we are primarily interested in string probes, we will focus on the ten-dimensional string metric  (\ref{stringmet}).  It is also convenient to factor out the scale, $\sqrt{Q_1 Q_5}$,  from the metric and work with the dimensionless metric:
\begin{equation}
d \tilde s^2_{10} ~\equiv~ \frac{1}{\sqrt{Q_1 Q_5}}\,d s^2_{10} ~=~ \Pi  \, \bigg(  \,{d\tilde s}_6^2 ~+~ \frac{1}{Q_5} \ d \hat{s}^2_{4} \bigg)  \,.
  \label{stringmetdimless}
\end{equation}
Indeed, since the $\IT^4$ directions are completely orthogonal to the six-dimensional metric, we can factor out the $\IT^4$  and simply compute and examine $\cal A$ for the six-dimensional metric:
\begin{equation}
 \Pi  \,   {d\tilde s}_6^2   \,.
  \label{sixDstringmet}
\end{equation}
We will also simplify the angular dependence by defining:
\begin{equation}
\chi ~\equiv~ 2\, n  \, \psi  +   2\, \varphi_1   \,.
  \label{chidefn}
\end{equation}
As noted in the discussion of  (\ref{btrans}), when   $b$ is only slightly greater than $a$, the counter-rotating geodesics are localized in the cap, where the geometry is fairly smooth and close to that of global AdS$_3$.     We therefore focus  on the co-rotating ``BPS geodesics.''

There are two families of such geodesics, corresponding to the two possible choices of sign in (\ref{simpsols1}) and the two sets of velocities in (\ref{ssvels1}).  Amusingly enough, the first family of geodesics is ``too BPS;'' it has $\frac{d\psi}{d \lambda} =  \frac{d\varphi_1}{d \lambda} =0$ and so has $\frac{d\chi}{d \lambda} =0$.  It simply does not traverse the bumps in the metric at all!  Indeed, one finds that for these geodesics, $\cA \equiv 2 m^2$ for all values or $r$, $a$ and $b$. This is precisely the value for global AdS\, and so these geodesics do not sense any deviation from AdS created by the momentum charge and momentum wave in the superstratum.
  
 The second set of BPS geodesics results in non-trivial tidal forces because they have non-trivial motion in the $\varphi_1$-direction.  One can explicitly simplify $\cA$ to obtain:
\begin{equation}
\begin{aligned}
\cA  ~=~   m^2 \, \big(1 -&  (1-G(r)) \, \cos \chi \big)^{-2} 
 \Big( \, 2 + (1-G(r))^2 (1+9 G(r)^2) \\  
 &   + 4\, (1-G(r)) (1+2 G(r)^2) \, \cos \chi+ (1-G(r))^3 (1+ G(r)) \, \cos 2\,\chi  \Big)  \,.
\end{aligned}
  \label{cAres}
\end{equation}

\begin{figure}[t]
\centering
\includegraphics[width=0.45\textwidth]{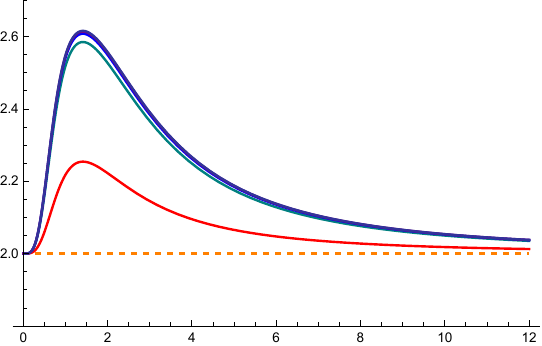} \qquad
\includegraphics[width=0.45\textwidth]{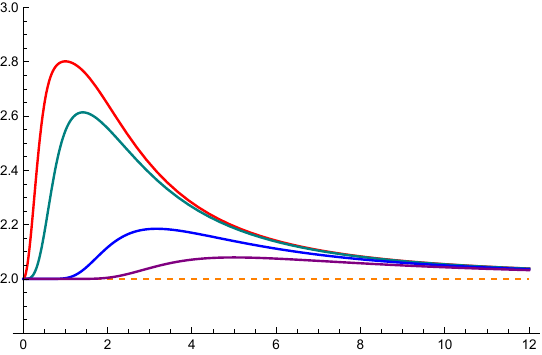} 
\caption{Plots  of the magnitude, ${\cal A}$, of the tidal force in the dimensionless metric (\ref{stringmetdimless}).  The plot on the left shows ${\cal A}$ as a function of $r$, for $a=1$, $n=1$, $\chi = \frac{\pi}{4}$, for $b = 1, 5 ,10, 20, 500$.  The higher curves correspond to higher values of $b$, and the curve does not change significantly for   $b \ge 10$.  The curve on the right shows ${\cal A}$ as a function of $r$, for $a=1$, $b=20$, $\chi = \frac{\pi}{4}$, for $n = 1, 2 ,10, 25$.  As $n$ increases, the curve flattens and the peak moves to the right.  The horizontal dashed line shows ${\cal A}$ for the AdS metric with $b=0$.}
\label{fig:Amag}
\end{figure}

\begin{figure}[t]
\centering
\includegraphics[width=0.5\textwidth]{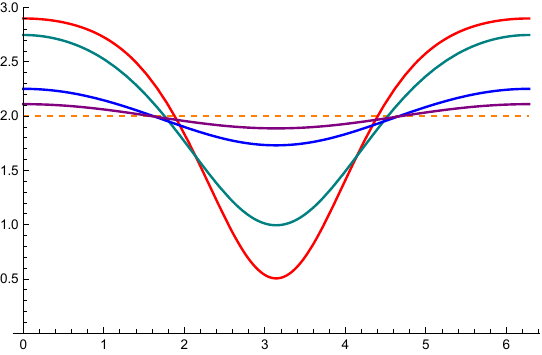} 
\caption{Plots  of the magnitude, ${\cal A}$, of the tidal force in the dimensionless metric (\ref{stringmetdimless}) as a function of $\chi$.  These plots are taken near the peak amplitude of ${\cal A}$, with $r= \sqrt{n}\, a$,  for $a=1$, $b=20$, for $n = 1, 2 ,5, 25$.  The higher amplitudes correspond to smaller values of $n$.   The horizontal dashed line shows ${\cal A}$ for the AdS metric with $b=0$.}
\label{fig:Avar}
\end{figure}

\begin{figure}[t]
\centering
\includegraphics[width=0.45\textwidth]{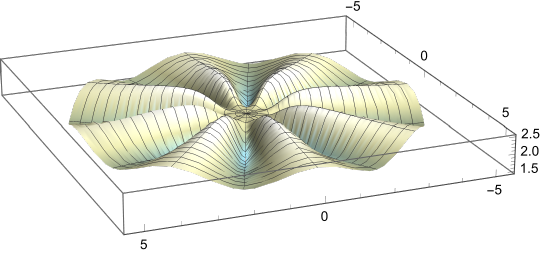} 
\includegraphics[width=0.45\textwidth]{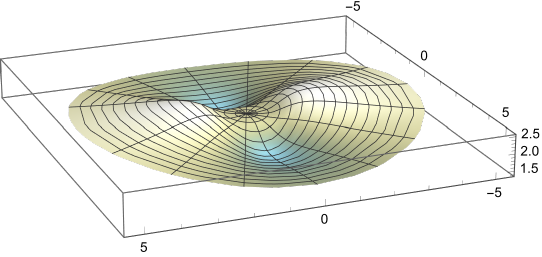} 
\caption{The magnitude of ${\cal A}$ in the BTZ throat and cap as a function of $(x_1,x_2) = r\,(\cos \psi\,, \sin \psi)$, on the left, and  as a function of $(x_1,x_2) = r\,(\cos \varphi_1\,, \sin \varphi_1)$, on the right.   This geometry has $a=1$, $n=4$ and $b=10$. Since $n=4$, there are thus four times as many bumps in the function on the left as there are in the function on the right.  The center shows the smooth AdS cap, and the folds reveal the bumps at the junction zone between the throat and the cap. }
\label{fig:bumps}
\end{figure}

To get a sense of the structure of this function we have plotted its features in Figs.~\ref{fig:Amag}, \ref{fig:Avar} and~\ref{fig:bumps}.   In general, the peak value of ${\cal A}$ increases with $b$, but rapidly saturates at large  $b$ (see Fig.~\ref{fig:Amag}) at a value that is not much larger  (less than 50\% higher) than the AdS tidal force.  The second plot in Fig.~\ref{fig:Amag} shows that if one increases $n$, the peak tidal force actually decreases and the peak flattens and moves to larger values of $r$.  For $ n>20$, the tidal forces are very close to their AdS values.

The tidal forces, and hence ${\cal A}$,  have non-trivial dependence on $\chi$.   Fig.~\ref{fig:Avar} shows that the fluctuations are largest for $n=1$, and their amplitudes decrease with $n$. The amplitude  of the fluctuations in ${\cal A}$ is of the same order as the  AdS value of ${\cal A}$.

Our {\it a priori} expectations were that the amplitudes of the tidal forces would be large and increase with $n$.  We anticipated the tidal forces depicted in the first diagram in Fig.~\ref{fig:bumps}:  an extremely bumpy function near the transition zone, and we thought that a null geodesic would bounce across this at ultra-relativistic speeds creating the large tidal forces.   Ironically, the non-BPS geodesics cannot enter this bumpy region, and one of the BPS geodesics does not traverse the bumps at all, while the other only traverses the bumps on the $S^3$.  Thus the only effect of $n$ on the tidal forces comes from the amplitude of $G(r)$, and this decreases with $n$.   While the first plot in Fig.~\ref{fig:bumps} depicts the tidal bumps the the spatial, $(r,\psi)$ directions, of the geometry of $\cK$, it is Fig.~\ref{fig:Avar} and  the second plot in  Fig.~\ref{fig:bumps} that best represents the tidal forces as a result of orbiting in $\varphi_1$ and not $\psi$.

Thus, unlike infalling geodesics, the geometry and the motion of the probe do not amplify the tidal forces.   These forces fluctuate  by amounts of $\cO(1)$ in units set by the AdS radius. We now consider what this can do to stringy probes.

\subsection{Stringy excitations}
\label{ss:StringyEffects}

The string world-sheet action in the pp-wave limit,   (\ref{BPenrose1}), becomes 
\begin{equation}
S = \int d^2\sigma  \, \Big[\, 2\, (\partial_\alpha u)  \, (\partial^\alpha v) + \partial_\alpha z^i  \,\partial^\alpha z^j \delta_{ij} ~+~  A_{ab}(u)\, z^a z^b \, (\partial_\alpha u)\,( \partial^\alpha u) ~+~ \ldots\Big] \, ,
  \label{ppaction}
\end{equation}
where  $\dots$ represent additional terms arising from the NS $B_{\mu \nu}$ field and the fermionic fields. Since we are looking for a qualitative understanding, we will omit these terms in the following discussion. 

In the light-cone gauge one takes:
\begin{equation}
u = \alpha' E  \, \sigma_0 \, ,
\label{lcgauge}
\end{equation}
and then the the $k^{\text{th}}$ oscillator mode satisfies:
\begin{equation}
\partial_{\sigma_0}^2 z^a + k^2 z^a + (\alpha' E)^2 A^a_{~b}(u)z^b = 0 \, .
\end{equation}

Generally, the matrix $A^a_{~b}$ is non-diagonal and thus generates a rotation of $z^i$.   As depicted in Fig.~\ref{fig:Avar}, the matrix  exhibits oscillatory behavior, and this will be reflected in its eigenvalues.  The actual behavior is a rather complicated function, but to give a broad-brush description of string excitations, we focus on a single eigenvalue, $\hat A$, and approximate its form by:
 \begin{equation}
 \hat A \sim \alpha_0  + \alpha_1\,  n^{-s} \cos (\alpha_2 \, \alpha' E\sigma_0) \, .
 \end{equation}
 where, based on Fig.~\ref{fig:Avar} and the velocities depicted in Fig.~\ref{fig:co-geodesics1} for the second family of BPS geodesics, the parameters  $\alpha_0$, $\alpha_1$ and  $\alpha_2$ are numbers of order unity, and $s > 0$ is chosen to match the diminishing amplitude of the eigenvalues with $n$. 
The primary contribution  to $\alpha_0$ comes from the constant curvature of the unperturbed ${\rm AdS}$ metric.  To arrive at the oscillatory term, recall that the plot in   Fig.~\ref{fig:Avar} is in terms of $\chi$, defined in (\ref{chidefn}), and the velocities are all of $\cO(1)$ in their dependence on $\varphi_1$, or $u= \alpha' E \sigma_0$.  Thus we consider the following mode equation
\begin{equation}
\Big[\partial_{\sigma_0}^2  + k^2  + \alpha_0 (\alpha' E)^2 + \alpha_1 ( \alpha' E)^2 n^{-s} \cos (\alpha_2 \, \alpha' E\sigma_0)\Big] z = 0 \, .
\label{ppsq}
\end{equation}

It is also important to recall that in analyzing the tidal forces we have factored the length scale, $(Q_1 Q_5)^\frac{1}{4}$,  out the metric and are working with the dimensionless metrics  (\ref{confsixmet}) and  (\ref{simpmet}).   The tidal tensor has dimensions of $L^{-2}$, and so  $\alpha_0$ and $\alpha_1$ are actually numbers of order unity times $(Q_1 Q_5)^{-\frac{1}{2}}$, and  $\alpha_2$ is a number of order unity times $(Q_1 Q_5)^{-\frac{1}{4}}$.  The differential operator in (\ref{ppsq}) is then dimensionless because $[\alpha'] \sim L^{2}$, the center of mass energy, $E$, has dimensions $[E] \sim L^{-1}$. 
It is also useful to recall that (see, for example, \cite{Martinec:2020cml})  that one has:
\begin{equation}
\label{eq:quantizedcharges}
Q_1   \,Q_5   ~=~   \frac{(2\pi)^4\,g_s^2 \,\alpha'^4}{V_4} \,  n_1 \, n_5  \,,
\end{equation}
where $n_1$ and  $n_5$  are the quantized charges and $V_4$ is the volume of the four-dimensional compactification torus.

The dynamics is therefore entirely controlled by the value of 
\begin{equation}
\nu ~\equiv~  \frac{\alpha' E}{(Q_1 Q_5)^{\frac{1}{4}}} ~\equiv~   \frac{V_4^{\frac{1}{4}} \, E }{2\pi\,g_s^{\frac{1}{2}}  (n_1 n_5)^{\frac{1}{4}}}   \, .
\label{contparam}
\end{equation}
For $\nu  \ll 1$, the amplitude of the perturbations is  extremely small  and the frequency of the perturbation, $\alpha' E$, is much less than the lowest mode of  the string, and so the string  will remain in its ground state.  

Stringy excitations will only become pronounced for  $\nu  \gtrsim 1$.   In this range the energy of the probe is large but it remains within the probe approximation.  Indeed, from the perspective of asymptotically-flat superstrata,  a significant gravitational back-reaction will arise when the energy, E, is comparable with the mass of the background, which would mean:
\begin{equation}
E  ~\sim~   G_5^{-1} \, (Q_1 + Q_5) ~\sim~  \frac{ 8\, V_4 \, R_y}{g_s^2 \,\alpha'{}^4}\, (Q_1 + Q_5)    \, .
\label{Ebackreact}
\end{equation}
From the AdS/CFT perspective, a heavy operator has $E \sim c \sim Q_1 Q_5$.  Thus, however one slices it,
\begin{equation}
E  ~\sim~   \alpha'{}^{-1} \, (Q_1 Q_5)^{\frac{1}{4}} ~\sim~  \frac{2\pi \,g_s^\frac{1}{2} }{V_4^\frac{1}{4}} \,  ( n_1 \, n_5)^\frac{1}{4}
\label{Escrambling}
\end{equation}
remains well within the bounds of the probe approximation.  

On the other hand, this energy is a large multiple of the Kaluza-Klein energy, and so such a particle will have many channels in which it can interact with the background.  Nevertheless, it is interesting and instructive to examine how such a string probe can become excited as a result of its motion through the  background metric of the superstratum.  Indeed, one can easily see how such interactions can grow exponentially and could therefore lead to chaotic excitations of the string.

\subsection{Fluctuating tidal forces }
\label{ss:resonances}

Motion in periodic potentials has a long history in physics, and most particularly with Bloch wave solutions to the Schr\"odinger equation.  The string dynamics involves a  second-order linear differential equation with a periodic ``potential'' term and represents a classic situation in Floquet theory.  One can always write the equations as a first-order system:
\begin{equation}
\frac{d}{dt} \, X(t)  ~=~ A(t)\, X(t) \,,
  \label{system1}
\end{equation}
where $X(t)$ is a $n$-dimensional vector and $A(t+T) = A(t)$ is an $n \times n$ matrix with period $T$.  The solutions to this system have the form 
\begin{equation}
X(t)  ~=~e^{i \mu \,t }\, P(t) \,,
  \label{Floquet}
\end{equation}
where $P(t+T) = P(t)$ is a periodic vector function and $\mu \in \IC$ is a complex number.   There are $n$ independent such solutions to this system with generically $n$ different values, $\mu_j$,   for $\mu$.  The $\mu_j$  measure the failure of periodicity and are known as the characteristic, or Floquet,  exponents\footnote{We have added an extra factor of $i$ in the exponent  in (\ref{Floquet}) so as to make the conventions consistent with the standard conventions for Mathieu functions.}. The imaginary part of $\mu_j$ is the Lyupanov exponent of the solution, and if it is  negative, the solution grows exponentially with time.  In this way, periodic tidal forces can lead to chaotic scrambling of strings.

It is entirely possible that the solution to the system has vanishing exponents, or simply ${\rm Im}(\mu_j) \ge 0$, so that the solutions do not grow in time. This, of course, depends on the details of $A(t)$.  The tidal tensors we are considering are already very complicated and the tidal tensors for more general superstrata are going to be even more complex.  Therefore, rather than making an exhaustive search for instabilities in the system we have here, we will  make some general observations about a much simpler system whose features resemble the tidal system of interest.

The Mathieu equation 
\begin{equation}
\frac{d^2}{dt^2} Y(t) ~+~ \big( a ~-~ 2\, q\, \cos (2t) \big)\, Y(t) ~=~0 \,,
  \label{Mathieu}
\end{equation}
is well studied and understood, and  has a potential that is qualitatively similar to those depicted in Fig.~\ref{fig:Avar}.  The equation, and hence the solution space, is invariant under $t \to -t$, which means that any imaginary exponent signals an instability.  It is also worth noting that the sign of $q$ can be flipped by sending $t \to t+ \frac{\pi}{2}$ and so stability issues are invariant under $q \to -q$.   For $q=0$ the equation is that of the simple harmonic oscillator and the frequency is $\sqrt{a}$.

\begin{figure}[t]
\centering
{\includegraphics[width=0.45\textwidth]{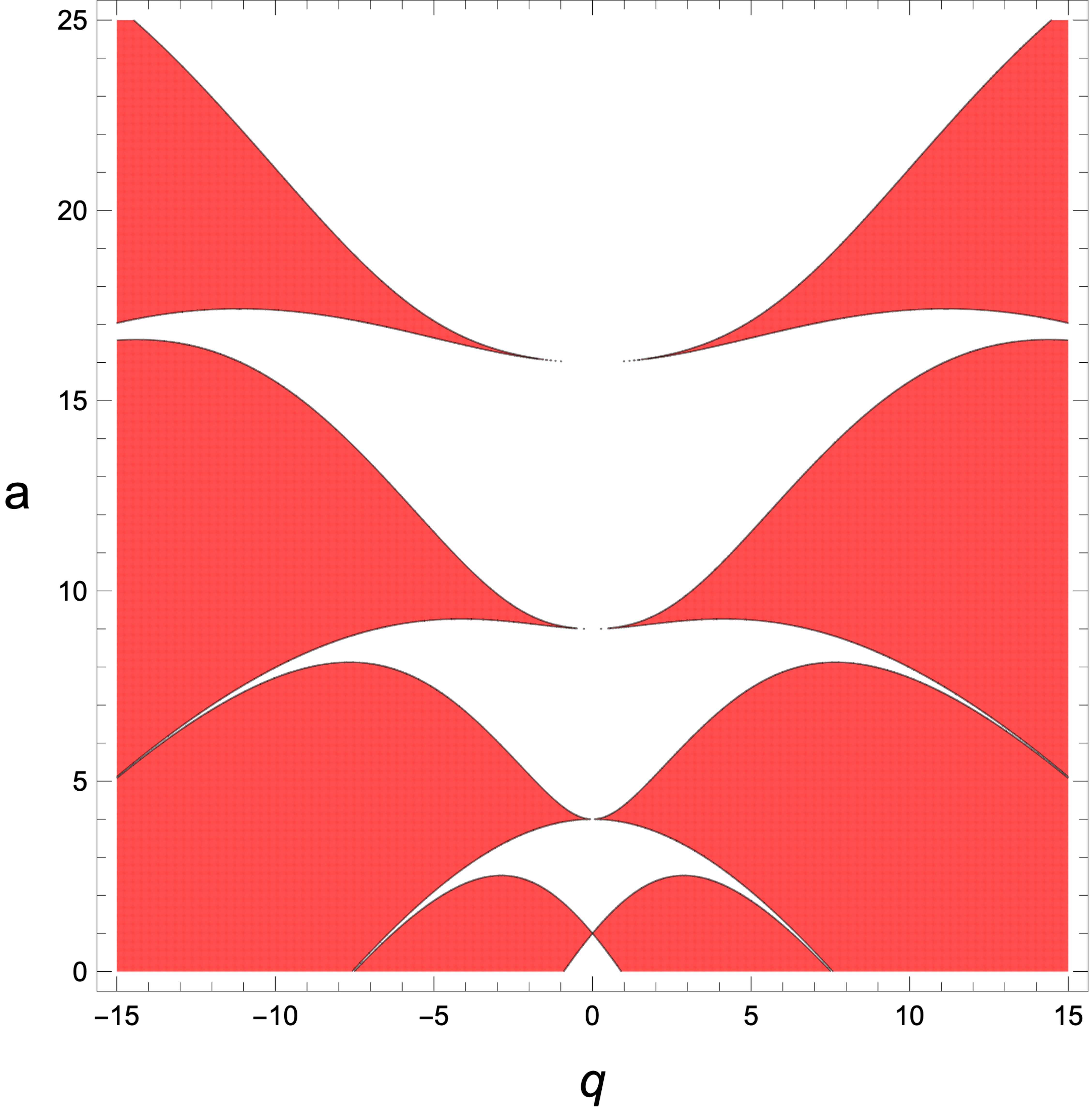}}\qquad\qquad
   \vbox{
    \hbox{ \includegraphics[width=0.4\textwidth]{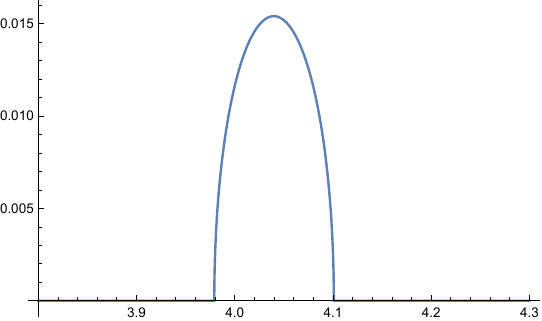}} 
    \bigskip
    \hbox{ \includegraphics[width=0.4\textwidth]{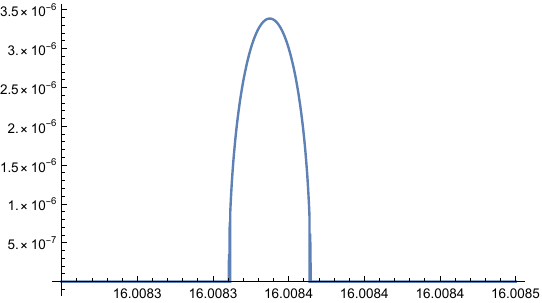}}}
\caption{ The left-hand plot shows the regions of stability (white) and instability (red) in the $(q,a)$ parameter space of the Mathieu equation. The instability regions extend all the way to the $q=0$ axis, meeting it at  $(q=0, a = j^2)$ for $j \in \ZZ_+$. (The apparent gaps in the plot as one approaches $q=0$ are artifacts of the numerical resolution.) The plots on the right show the unstable Lyupanov exponents for $q= \frac{1}{2}$ and  $ a \approx 4$ and $a \approx 16$. As $a$ increases, the instability bands get narrower and the exponents get smaller. }
\label{fig:stability}
\end{figure}		

The exponent, $\mu$, for the Mathieu equation is a well-studied function of $(q,a)$ and so it is straightforward to map out the regions of stability and instability in this parameter space. We have plotted these in Fig.~\ref{fig:stability}.  At large $|q|$ the solutions are largely unstable, with very small islands of stability for specific, narrow ranges of $a$.  For $q=0$, the solutions are harmonic and thus stable, or marginally stable.  The instability regions are curvilinear triangles, each with  an apex at $(q=0, a = j^2)$ for $j \in \ZZ_+$.  The regions of instability reach all the way down to $q=0$, however, for larger values of $a$ the instability zone for small $q$ is extremely narrow and the Lyupanov exponent is very small.  In  Fig.~\ref{fig:stability} we have also plotted the imaginary parts of the exponent in the regions $a \approx 4$ and $a \approx 16$ with $q =\frac{1}{2}$.  These figures illustrate how narrow the instability region is, and how small the Lyupanov exponent becomes.  

One of the interesting features of these Floquet instabilities is that they are far more extensive and powerful than the much simpler phenomenon of resonant forcing.   The fact that the potential function is periodic generates instabilities that grow {\it exponentially} in a region around {\it every integer multiple} of the fundamental frequency of the underlying potential, and the larger the value of $q$, the wider the instability range in $a$.  Thus, in principle, even a low-energy string probe encountering small bumps in the geometry can excite an exponentially growing string-scale resonance in the probe.  This requires fine tuning and the growth will be slow, but it will  be non-zero and exponential.  For higher-energy probes, the resonance range and exponents will be larger.  When $q$ becomes comparable with $a$, the instabilities regions are dominant.

\section{Conclusions}
\label{sec:Conc}

We have analyzed circular geodesics in a superstratum geometry and this has revealed some interesting features. 

First, the co-rotating (relative to the superstratum momentum) orbits behave as one would expect of BPS particles:  they can be located at any radial distance from the center of the superstratum. They can thus be used as probes of any region within the geometry. Conversely,  the counter-rotating orbits have a strongly non-BPS behavior and, even for relatively small superstratum momentum charge, these orbits become deeply bound in the geometry and quickly localize in the cap.  

We have shown that the counter-rotating orbits are a surprisingly accurate bellwether of the black-hole bound.   Some microstate geometries, and some superstrata in particular, are ``over-rotating:'' they have a sufficiently large angular momentum so that the corresponding black-hole would have a naked singularity and no horizon.  Adding charge, and hence mass, to such a microstate geometry can take the geometry across the ``black-hole bound,'' where the corresponding black-hole  develops a horizon, and the area of the horizon grows as more mass is added.    As the black-hole bound is crossed, the   superstratum itself seems rather oblivious to the transition: the geometry  does not change  significantly, remaining smooth and horizonless as the black-hole-like throat becomes incrementally deeper.  As we have seen in this paper, there is, however, a dramatic change in the permissible location of counter-rotating geodesics: {\it exactly} when  one crosses the black-hole bound, the counter-rotating geodesics  become bound into the geometry, and after a very small increase beyond the bound, the geodesics are trapped in the cap at the bottom of the superstratum.

While we have only considered circular geodesics here, we expect  similar results will hold for non-circular bound orbits, at least in the neighborhood of our circular orbits.

Most of the earlier studies of tidal forces on geodesic probes of microstate geometries have focussed on infalling particles  \cite{Tyukov:2017uig, Bena:2018mpb, Bena:2020iyw,Martinec:2020cml,Ceplak:2021kgl}.  These probes are scrambled by string-scale tidal forces because the very long throat of the microstate geometry behaves like a particle accelerator and the non-trivial multipoles sourced by the cap on the geometry are then hugely magnified by the extremely high energies attained by the probes during infall.   Indeed, if an infalling massless string starts with a very  low-energy, it will still become highly excited by the time they reach and pass through the cap \cite{Martinec:2020cml,Ceplak:2021kgl}.  This results in ``tidal trapping'' and scrambling of such probes.  

Motivated by the ideas of long-term trapping near evanescent ergosurfaces, this paper examined probes that are already trapped in the geometry, considering the strength of the trapping and the scrambling process.  While there is no longer the huge amplification of tidal effects caused by the blue-shift of infall, we had originally expected strong scrambling effects from the ``corrugations'' in the geometry.  This is because a massless string travels at the speed of light and so would encounter the bumps in the geometry, especially around the transition region between the throat and the cap,  at extreme speed. We had thought that  this might result in a slightly smoothed analog of the ``null shockwave'' that a probe is expected to  encounter when there is structure at the horizon of a black hole or black ring \cite{Bena:2004td,Horowitz:2004je,Bena:2014rea}.  

This is not what we found.  First, when the superstratum has a significant momentum charge, and hence a very long throat, the counter-rotating geodesics are all trapped in the cap and cannot explore the corrugations. The co-rotating geodesics come in two families: one is co-moving with {\it all the corrugations} in  the space-time, $\cK$, and on the $S^3$.  The other family is co-moving with the corrugations in $\cK$, but traverses the corrugations on the $S^3$.   We have referred to these as ``BPS geodesics'' because they resemble floating branes \cite{Bena:2009fi}  in that both families sit at fixed $(r, \psi)$ in $\cK$, apparently feeling no force.  This suggests that they could be probe representatives of BPS excitations of the system. Indeed, it is natural to conjecture that the family that is also fixed on the $S^3$ is simply a probe representative of known superstrata excitations.  The other family is a little more surprising because it moves on the $S^3$.   This suggests that there may be new excitations that are not BPS in the full six-dimensional geometry but could be BPS from the perspective of three-dimensional supergravity discussed in \cite{Mayerson:2020tcl,Ganchev:2021pgs,Ganchev:2021iwy,Ganchev:2022exf,Ganchev:2023sth}.

The tidal forces encountered by the first family are unmodified from those of global AdS$_3$, while the second family encounters oscillating tidal forces of the same scale as those of a global AdS$_3$ geometry.  It is possible that the small tidal forces encountered by probe is an artifact of the particular single mode superstratum that we used here:  The massless string indeed travels at the speed of light but in our example, this motion is  entirely on the $S^3$, on which the corrugations are very slowly varying.  Corrugations with a high mode number on both the $S^3$ and the AdS$_3$ could result in larger tidal forces, however the amplitudes of the corrugations will decrease as mode numbers increase.

Even though the tidal forces remain relatively modest, they can still become large enough, within the range of the probe approximation, to excite a string. The corrugations in the metric mean that the string probe develops  a periodic mass term and this generically produces a much stronger effect than the mere resonance of periodic force.  A periodic mass term can result in exponential growth of the excitations, with Lyapunov exponents that depend on the amplitude of the variations and the proximity to a resonance with the string modes.  One thus expects the scrambling process to be chaotic.  

More broadly, it was originally expected that  for microstate geometries to reproduce the black-hole-like behavior of trapping and scrambling of matter, it would  require one to calculate back-reactions and compute energy transfers between probes and the background geometries and fluxes.  While these such back-reactions will be an essential part of the full story of microstate geometries, one of the remarkable surprises of the last few years is that one can see the onset of such trapping and scrambling entirely within a  probe approximation  \cite{Tyukov:2017uig, Bena:2018mpb, Bena:2020iyw,Martinec:2020cml,Ceplak:2021kgl,Berenstein:2023vtd}:  probes become trapped in the cores of microstate geometries and  scramble chaotically in quite a variety of ways, even when the background geometry is fixed to that of a single (coherent combination of) microstates.  The fully back-reacted story will be much richer and far more complicated, but it is very encouraging  that this essential black-hole-like behavior is already visible within the most basic of microstate geometries.



\section*{Acknowledgments}

\vspace{-2mm}
We are grateful for valuable conversations with Iosif Bena, Nejc \v{C}eplak, Bogdan Ganchev, Anthony Houppe and Rodolfo Russo.
The work of NPW is supported in part by the DOE grant DE-SC0011687. The work of BG and NPW is supported in part by the ERC Grant 787320 - QBH Structure.  The work of SDH is supported by ERC Grant 787320 - QBH Structure and by KIAS Grant PG096301.

\newpage

\begin{adjustwidth}{-1mm}{-1mm} 

\bibliographystyle{utphys}      

\bibliography{references}       

\end{adjustwidth}


\end{document}